\documentclass[pra,amsmath,amssymb,twocolumn]{revtex4-1}

\usepackage{graphicx}
\usepackage{dcolumn}
\usepackage{bm}
\usepackage[utf8]{inputenc}
\usepackage[english]{babel}
\usepackage{amsfonts}
\usepackage{hyperref}
\usepackage{physics}
\usepackage{epstopdf}
\usepackage{mathtools}
\usepackage[capitalise]{cleveref}
\usepackage{float}
\usepackage[space]{grffile}
\usepackage{color}
\usepackage[normalem]{ulem}
\usepackage{subfigure}

\binoppenalty=\maxdimen
\relpenalty=\maxdimen

\addto\captionsenglish{}

\begin{document}

\title{Simple variational ground state and pure cat state generation in the quantum Rabi model}

\author{C. Leroux}
\author{L. C. G. Govia}
\author{A. A. Clerk}
\affiliation{Department of Physics, McGill University, Montr\'eal, Qu\'ebec, Canada.}

\begin{abstract}
We introduce a simple, physically-motivated variational ground state for the quantum Rabi model, and demonstrate that it provides a high-fidelity approximation of the true ground state in all parameter regimes (including intermediate and strong coupling
regimes).  Our variational state is constructed using Gaussian cavity states and nonorthogonal qubit pointer states, and contains only three variational parameters.  We use our state to develop a heuristic understanding of how the ground state evolves with increasing coupling, and find a previously unexplored regime where the ground state corresponds to the cavity being in a nearly pure Schr\"odinger cat state.
\end{abstract}

\maketitle


\section{Introduction}

The quantum Rabi model is a well established paradigm of light-matter interaction \cite{Rabi:1936aa,Rabi:1937aa,Braak:2016aa}. It describes a two-level atom or qubit interacting coherently with an electromagnetic cavity, with a coupling strength strong enough that the rotating-wave approximation cannot be applied; as such, the system is no longer accurately described by the simpler Jaynes-Cummings model \cite{Jaynes63}. The so called ultrastrong coupling regime necessary to observe quantum Rabi physics has been achieved experimentally \cite{Anappara:2009aa,Gunter:2009aa,Niemczyk:2010aa,Forn-Diaz:2010aa,Yoshihara:2016aa}, and recently new approaches have been implemented to simulate the quantum Rabi model \cite{Langford:aa}.

Despite renewed experimental interest and a long history of theoretical study, exact analytical solutions to the quantum Rabi model have only recently been discovered \cite{2011Braak,Chen:2012aa,2013Lee}. The complexity of these analytical solutions implies that approximate descriptions are still of great utility, as they provide a potential means for understanding the physics of the model more intuitively \cite{Irish:2007aa,Casanova:2010aa,2010Nori,2010Choi,2012Solano,Liu:2015aa}. In particular, the properties of the eigenstates, especially the ground state, must be well understood for applications to quantum information, including metrology and error correction \cite{1999Munro,2001Hirota,2004Milburn,Cheng:2014aa}, or for extensions to many-body systems \cite{Schiro:2012aa,Schiro:2013aa}, and these properties are more easily understood using approximate solutions \cite{Chen:1989aa,Shore:1973aa,Stolze:1990aa,Irish:2007aa,2010Choi}.

In this paper we propose a simple variational ground state for the quantum Rabi model, one which approximates the true ground state with extremely high fidelity for essentially all parameters.
Our variational state has a very simple form, depending on only three free parameters, and only uses Gaussian cavity states. It exactly recovers the true ground state in both the weak and infinite coupling limits, and has very high fidelity with the exact ground state even in intermediate coupling regimes.
As our variational state contains only Gaussian cavity states, its success is quite remarkable, given that the cavity inherits a strong effective qubit-induced nonlinearity in the ultrastrong coupling regime.

\begin{figure}[t]
  \includegraphics[width = 0.2\textwidth]{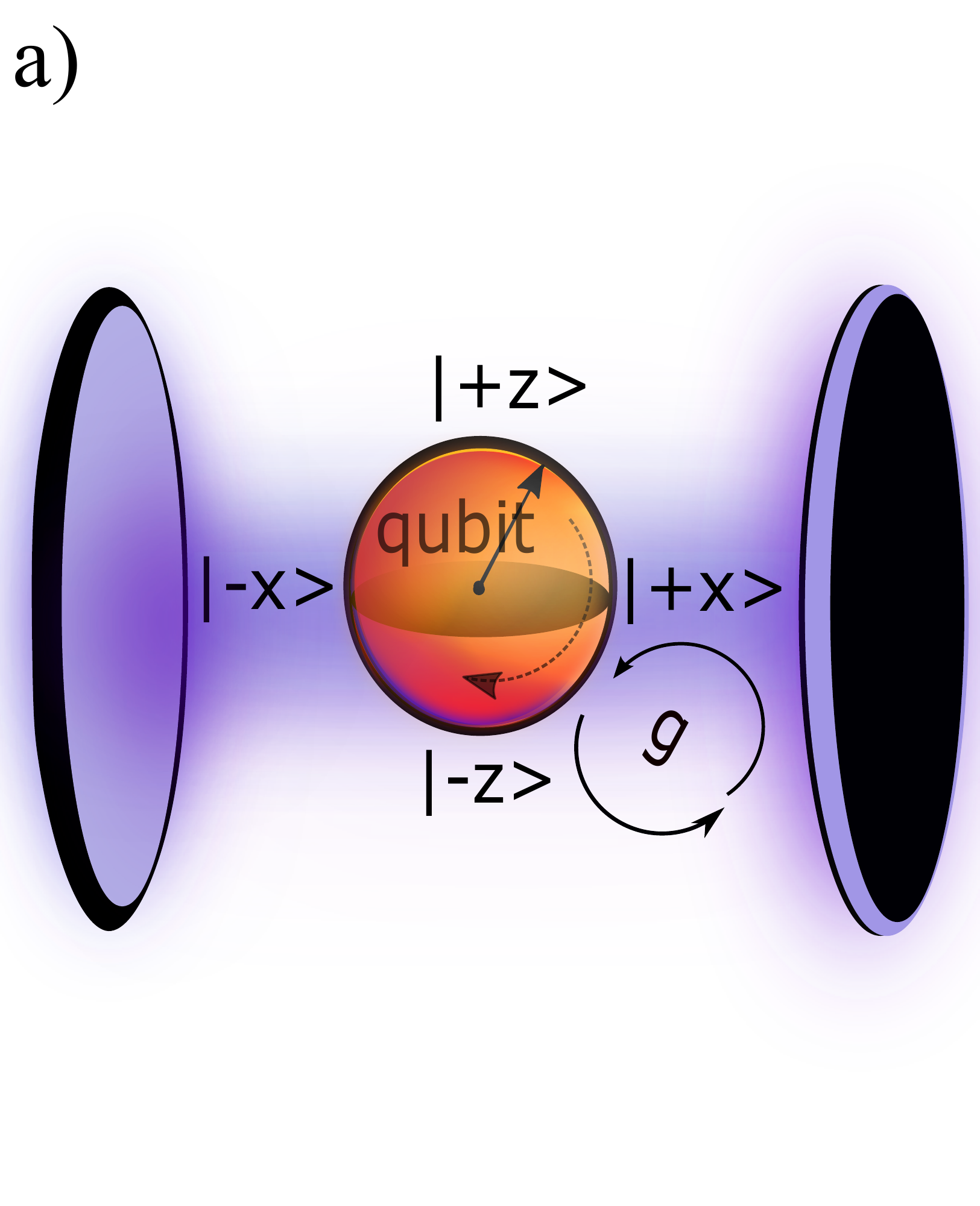}
  \includegraphics[width=0.25\textwidth]{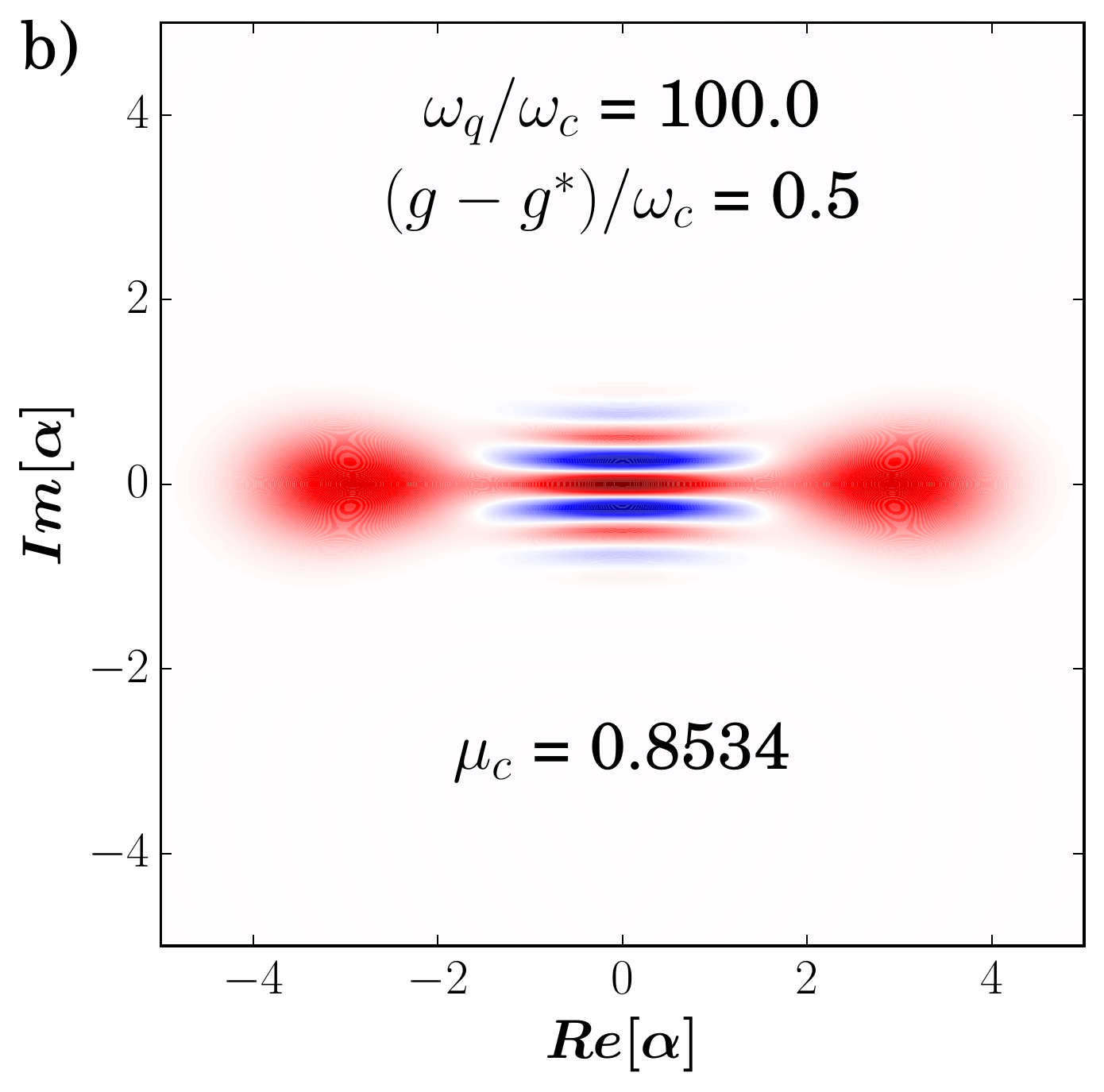}
  \caption{\textbf{(a)} Schematic of the interaction between an electromagnetic cavity and a two-level atom, as described by the Rabi Hamiltonian.
  \textbf{(b)} Example cavity Wigner function corresponding to the Rabi model ground state in a parameter regime where the cavity is in a relatively pure state;
  parameters indicated in the caption, with $g^*$ denoting the cross-over coupling defined in \cref{eq:gstar}, and $\mu_c$ denoting the purity of the reduced cavity state.  The variational ground state described in this paper provides insight into when and how such pure cat states can be formed.}
  \label{fig:scheme}
\end{figure}

The accuracy of our variational state in capturing the true Rabi model ground state means that it can be used to both qualitatively and quantitatively understand the physics of the ultrastrong coupling regime.  Here, we use it to explore a phenomena that has not received much attention previously: the possibility of having ground states that correspond to the cavity being in an almost pure Schr\"odinger cat state (c.f.~Fig.~\ref{fig:scheme}(b)). Such states are of interest to applications in quantum information \cite{1999Munro,2001Hirota,2004Milburn,Cheng:2014aa}. In this regime, the coupling is sufficient for the the qubit to mediate a strong cavity nonlinearity, leading to highly nonclassical Schr\"odinger cat states in the cavity.  At the same time however, the coupling strength is not strong enough to induce significant qubit-cavity entanglement.  The result is that the reduced cavity state is approximately a pure cat state.

The structure of the paper is as follows: in Sec.~\ref{sec:limits} we quickly review the quantum Rabi model and its solutions in known limits, while in Sec.~\ref{sec:GSAnsatz} we present our variational state for the ground state. In Sec.~\ref{sec:CatStates} we use our variational state to understand the formation of pure Schr\"odinger cat states in the cavity in the ultrastrong coupling regime, and in Sec.~\ref{sec:HigherLevels} we discuss extensions of our variational solution to the first excited state. Finally, we note that in certain regimes, our variational state is similar to that used by Hwang and Choi in Ref.~\cite{2010Choi}; there are however crucial differences in general. This is discussed in more detail in Appendix \ref{sec:Comp}, where we compare our variational state to that of Ref.~\cite{2010Choi}.


\begin{figure*}[t]
\centering
\includegraphics[width=0.90\textwidth]{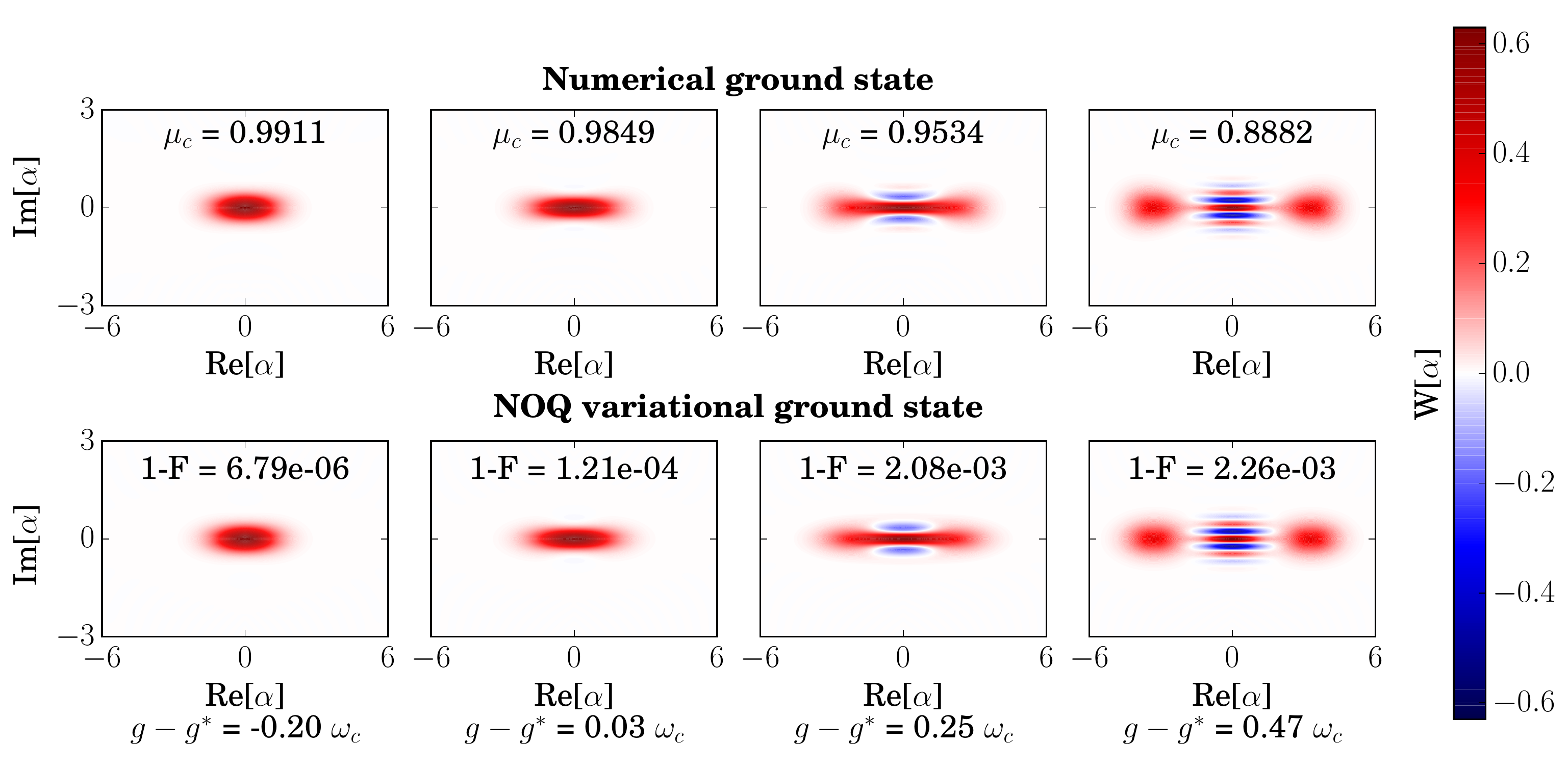}
\caption{Evolution of the cavity Wigner function in the ground state of the Rabi Hamiltonian of \cref{eq:RabiHamiltonian}, as a function of the coupling $g$ (measured from the approximate crossover scale $g^* \equiv \sqrt{\omega_c\omega_q}/2$).  The ratio $\omega_q/\omega_c$
of qubit and cavity frequencies was fixed at a large value $176.0$ to highlight the interesting regime (at ultrastrong couplings) where almost pure Schr\"odinger cat states are formed in the cavity. The top row shows the numerical ground state and the purity ($\mu_c$ defined in \cref{eq:purity}) of its reduced cavity state. The bottom row shows the optimized variational non-orthogonal qubit (NOQ) state and its fidelity ($F$ defined in \cref{eq:Overlap}) with the numerical state. The cavity Wigner function is defined by $W(\alpha)=2{\rm Tr}[\hat{\rho}_{\rm cav}\hat{D}(\alpha)\exp(i\pi\hat{a}^{\dagger}\hat{a})\hat{D}^{\dagger}(\alpha)]$ \cite{1969Cahill}, where $\hat{\rho}_{\rm cav}$ is the reduced density matrix of the cavity in the ground state.}
\label{fig:WignerPlots}
\end{figure*}

\section{Hamiltonian and known limiting cases}
\label{sec:limits}

The interaction between an electromagnetic cavity and a two-level system (qubit), shown in \cref{fig:scheme}(a), is described by the Rabi Hamiltonian ($\hbar = 1$)
\begin{equation}
\hat{H} = \omega_c\hat{a}^{\dagger}\hat{a} + g \hat{\sigma }_x \left(\hat{a}^{\dagger}+\hat{a}\right) + \frac{\omega_q}{2} \hat{\sigma}_z, \label{eq:RabiHamiltonian}
\end{equation}
where $\hat{\sigma }_x$ and $\hat{\sigma }_z$ are the Pauli operators for the qubit, and $\hat{a}$ $(\hat{a}^\dagger)$ is the lowering (raising) operator for the cavity. We are interested in the ground state of the Rabi Hamiltonian, and in \cref{fig:WignerPlots} Wigner functions for the reduced cavity ground state (found via numerical diagonalization) are shown for a range of qubit-cavity couplings. We have chosen to plot the Wigner functions for a large value of the ratio $\omega_q/\omega_c$ as this highlights the interesting regime with large Schr\"odinger cat states. However, the reduced cavity state for our variational solution (also shown in \cref{fig:WignerPlots}) has excellent agreement with the exact state for other (smaller or larger) values of $\omega_q/\omega_c$.

As can be seen, the ground state smoothly evolves from a weak coupling state where the cavity is squeezed (but minimally entangled with the qubit), to a qubit-entangled cat state, resulting in a mixture in the cavity. For interaction strengths in between these limiting cases, relatively pure cat states can be found in the cavity. To better understand the ground state's properties, and gain intuition into designing a variational state, we now examine the limiting cases of equation \cref{eq:RabiHamiltonian}.

\subsection{Large Coupling Limit}

In the limit $\omega_q/g \rightarrow 0$ we can ignore the last, qubit-only term in \cref{eq:RabiHamiltonian}. In this case $\hat{\sigma}_x$ becomes a conserved quantity, and the qubit-cavity interaction acts as a static $\hat{\sigma}_x$-dependent linear force on the cavity. This force drives the cavity into one of two coherent states $\pm \alpha$ (depending on the value of $\hat{\sigma}_x$), conditioned on the state of the qubit in the $\hat{\sigma}_x$ basis. In this case, a polaron transformation via the unitary
\begin{equation}
  \hat{U}_{\rm P} = \ketbra{+x}{+x}\hat{D}(\alpha) + \ketbra{-x}{-x}\hat{D}(-\alpha),
\end{equation}
diagonalizes the Hamiltonian, where $\hat{D}(\alpha) = \exp\left(\alpha\hat{a}^\dagger - \alpha^*\hat{a}\right)$ is the usual displacement operator with $\alpha = -g/\omega_c$, and $\ket{\pm x}$ are the eigenstates of $\hat{\sigma}_x$.  The eigenstates are easily found in the polaron frame; transforming
back to the original lab frame, they have the form of displaced Fock states
\begin{equation}
  \ket{\psi_{\pm,n}^{\rm P}} = \hat{D}(\pm\alpha)\ket{n}\ket{\pm x},
\end{equation}
with the direction of the displacement determined by the state of the qubit. These states are two-fold degenerate, and the ground state manifold can also be described as a pair of fully entangled cat states
\begin{equation}
  \ket{\psi_{\pm}^{\rm ECS}} = \frac{1}{\sqrt{2}}\left(\ket{\alpha}\ket{x} \pm \ket{-\alpha}\ket{-x}\right), \label{eq:ECS}
\end{equation}
where $\ket{\alpha}$ denotes a coherent state of amplitude $\alpha$.

\subsubsection{Finite $\omega_q$}

We now turn to the case where $g/\omega_c \gg 0$ but $g/\omega_q$ is non-zero. Here, it is useful to make use of the fact that the Rabi Hamiltonian in \cref{eq:RabiHamiltonian} conserves the total number of qubit and cavity excitations modulo two, i.e. the parity of total excitation number, described by the operator $-\hat{\Pi}\hat{\sigma}_z$, is conserved. Here $\hat{\Pi} = (-1)^{\hat{n}}$ is the parity operator of the cavity alone. We thus transform to a frame where this conservation is explicit.  This is done via the unitary transformation
\begin{equation}
\hat{U}_{\Pi} = e^{i \frac{\pi}{4}(1-\hat{\Pi}) \hat{\sigma}_y},
\end{equation}
which rotates the qubit state if the cavity parity is odd. This results in the parity frame Hamiltonian
\begin{align}
\hat{H}_{\pm}  = \hat{U}_{\Pi}\hat{H}\hat{U}_{\Pi}^\dagger = \omega_c\hat{a}^{\dagger}\hat{a} - g \hat{\sigma}_z (\hat{a}^{\dagger}+\hat{a}) + \frac{1}{2} \omega_q \hat{\Pi} \hat{\sigma}_z. \label{eq:effectiveH}
\end{align}
In this new frame, $\hat{\sigma}_z$ is conserved, and corresponds to the conserved excitation parity of the original Hamiltonian. Parity transformations have been used extensively in the literature to understand the eigenstates of the Rabi Hamiltonian \cite{1987Messina,2010Choi,Casanova:2010aa,Wolf:2013aa}.  Here, we use the parity frame to to obtain intuition that helps motivate our variational state.

$\hat{H}_{\pm}$ is diagonal in the $\hat{\sigma}_z$ basis, and therefore we can write $\hat{H}_\pm$ as a direct sum of the Hamiltonians $\hat{H}_+$ and $\hat{H}_-$ which act on subspaces defined by $\left<\hat{\sigma}_z\right>= \pm 1$ respectively; we refer to these as parity subspaces. For nonzero $\omega_q$, the ground state will be in the parity subspace with $\left<\hat{\sigma}_z\right>= -1$
(i.e.~the subpsace where the total excitation number has even parity). We thus will focus on this subspace, with Hamiltonian
\begin{align}
  \hat{H}_{-}  = \omega_c\hat{a}^{\dagger}\hat{a} + g (\hat{a}^{\dagger}+\hat{a}) - \frac{1}{2} \omega_q \hat{\Pi}.
  \label{eq:ParityH-}
\end{align}
We consider the evolution of the ground state as $g$ is varied from $g\gg\omega_q$ to $g\sim\omega_q$.

$\hat{H}_-$ describes a cavity with a nonlinear interaction.  For $\omega_q = 0$, the nonlinearity vanishes, and the remaining
terms in $\hat{H}_-$ describe a displaced oscillator.  Thus, for $\omega_q=0$, the eigenstates of $\hat{H}_-$ are displaced Fock states, and the ground state is a simple coherent state $\ket{\alpha}$.  To return to the lab frame, we apply the unitary $\hat{U}_{\Pi}^\dagger$ to the parity frame state.  In practice, we simply decompose the parity-frame state into its even and odd photon number components $\ket{\psi_e}$ and $\ket{\psi_o}$.  For a state with total even parity, the state in the lab frame is then simply  $\ket{\psi_e} \ket{-z} + \ket{\psi_o} \ket{+z}$, where  $\ket{\pm z}$ are the eigenstates of $\hat{\sigma}_z$.  For the state
$\alpha$,   $\ket{\psi_{e/o}}$ are proportional to cat states, $\ket{\psi_{e/o}} \propto \left( \ket{\alpha} \pm \ket{-\alpha} \right)$.
The $\omega_q = 0$ ground state of $\hat{H}_-$ thus corresponds to the
fully entangled Schr\"odinger cat state $\ket{\psi^{\rm ECS}_{-}}$ defined in \cref{eq:ECS}.

For $\omega_q \neq 0$ the last term in $\hat{H}_-$ introduces an energetic preference for states with even photon-number parity.   Thus, as $\omega_q$ increases, we expect that the even-parity parts of the state $\ket{\alpha}$ are weighted more strongly than the odd-parity part of this state.
As such, we expect that the ground state in the parity frame can be described by a trial variational state of the form
\begin{align}
  \ket{\psi_-^{\Pi}[\alpha]} = \lambda_E\ket{\Phi_{+}[\alpha]} +  \lambda_O\ket{\Phi_{-}[\alpha]}, \label{eq:parans}
\end{align}
with $\lambda_E \ge \lambda_O$, where
\begin{align}
  \ket{\Phi_{\pm}[\alpha]} = \left(\ket{\alpha}\pm\ket{-\alpha}\right),
\end{align}
are unnormalized even and odd cat states, and $\alpha$ is a variational parameter.

To see what this state looks like in the original lab frame, we follow the procedure described above and apply $\hat{U}_{\Pi}^\dagger$ to the parity-frame state in \cref{eq:parans}.  We obtain then the lab-frame state
\begin{align}
&\ket{\psi_-} = \hat{U}_{\Pi}^\dagger\ket{\psi_-}_{\Pi} = \lambda_E\ket{\Phi_{+}}\ket{-z} +  \lambda_O\ket{\Phi_{-}}\ket{+z} \label{eq:ParECS} \\
\nonumber&= \ket{\alpha} \Big[ \lambda_E\ket{-z} + \lambda_O\ket{+z}\Big] +
	\ket{-\alpha} \Big[ \lambda_E\ket{-z} - \lambda_O\ket{+z} \Big].
\end{align}
Looking at the first line of \cref{eq:ParECS}, we see that if $\lambda_E \gg \lambda_O$, the reduced cavity state is a pure cat state. Thus, the energetic preference for a definite photon-number parity in the parity frame when $\omega_q > 0$ translates to a preference for pure cat states in the lab frame.  This behaviour is seen in \cref{fig:WignerPlots}.

In the lab frame, the preference for a definite photon number parity in Eq.~(\ref{eq:ParityH-}) has a direct implication for the qubit components of the ground state.  There are two qubit ``pointer states" in the second line of \cref{eq:ParECS}, one for each term.
Having a definite photon number parity results in $\lambda_E \neq \lambda_O$, and hence these qubit states are in general
nonorthogonal.  This nonorthogonality of course reduces the amount of qubit-cavity entanglement, and in the extreme limit can allow the cavity state to have a high purity.  The second line of  \cref{eq:ParECS} thus provides an important ingredient for building a variational ansatz: \emph{ one should use qubit pointer states that can be nonorthogonal from one another}.

To build a variational ground state that is effective for all parameter regimes, we will need to combine the above intuition with an aspect of the physics that is clearest by considering the weak coupling regime; we do this next.

\subsection{Weak Coupling Limit}

In the weak coupling limit $g/\omega_q \rightarrow 0$, we recover the non-interacting case where the Hamiltonian is trivially diagonalized by product states $\ket{n}\ket{\pm z}$, with the ground state given by $\ket{0}\ket{-z}$. For small but finite coupling, the interaction with the qubit generates a weak squeezing term for the cavity, leading to squeezing correlations in the ground state.  As this will also be an important ingredient in our variational ansatz, we discuss the origin of this squeezing in more detail.

Focusing on $g\ll\omega_q$, we can treat the interaction perturbatively.  We make a standard Schrieffer-Wolff transformation of the Rabi Hamiltonian in \cref{eq:RabiHamiltonian} to eliminate the interaction to leading order.  This corresponds to a unitary transformation $e^{\hat{A}}$ with generator
\begin{equation}
  \hat{A} = \frac{g\omega_c}{\omega_q^2-\omega_c^2}\left[\hat{\sigma }_x \left(\hat{a} - \hat{a}^\dagger\right) + i\frac{\omega_q}{\omega_c} \hat{\sigma }_y \left(\hat{a} + \hat{a}^\dagger\right) \right]. \label{eq:SWgen}
\end{equation}
Keeping terms up to second order, this results in the effective Hamiltonian
\begin{align}
e^{\hat{A}} \hat{H} \hat{e}^{-\hat{A}} \approx \omega_c\hat{a}^{\dagger}\hat{a}+\frac{g^2\omega_q\hat{\sigma}_z}{\omega_q^2-\omega_c^2} (\hat{a}+\hat{a}^{\dagger})^2 + \frac{\omega_q\hat{\sigma}_z}{2}. \label{eq:BSHam}
\end{align}
This is equivalent to deriving an effective Hamiltonian up to second order in perturbation theory.

Physically, the photon-number non-conserving terms in \cref{eq:BSHam} (which generate squeezing) arise from a second-order process involving both kinds of interaction terms present in \cref{eq:RabiHamiltonian}, $\hat{\sigma}_x\hat{a}^\dagger$ and $\hat{\sigma}_x\hat{a}$. Together, these terms can create or destroy pairs of photons, via a virtual state with a single photon and a flipped qubit state. As the qubit state remains unchanged by the net effect of this second order process, it is not energetically forbidden, even for $g \ll \omega_q$.

The ground state of \cref{eq:BSHam} is  $\hat{S}(-r)\ket{0}\ket{-z}$, where $\hat{S}(r) = \exp\left(r\hat{a}^{\dagger 2} - r^*\hat{a}^2\right)$ is the usual squeezing operator, and the squeeze parameter $r$ is real and given by
\begin{equation}
  e^{-2r} = \sqrt{1-\frac{4g^2\omega_q}{\omega_c(\omega_q^2-\omega_c^2)}}. \label{eq:sqzr}
\end{equation}

The approximate Hamiltonian in \cref{eq:BSHam} hits a parametric instability (i.e.~$e^{2r}$ diverges) when $g$ is increased to $g^*$, with
\begin{align}
g^* \equiv \frac{1}{2}\sqrt{\omega_c\omega_q\left[1-\left(\dfrac{\omega_c}{\omega_q}\right)^2\right]}.
\end{align}

The instability of  \cref{eq:BSHam} at large $g$ indicates a breakdown of our perturbative derivation of this Hamiltonian.  For $g \geq g^*$,
higher order terms will restore stability to the effective Hamiltonian, as the Rabi Hamiltonian is stable for all parameter regimes. Nevertheless, $g^*$ defines the onset of an effective cavity nonlinearity, and as a result is the boundary of the interesting regime where Schr\"odinger cat states are formed. In this case, we will often be interested in the large $\omega_q$ limit, and we will typically approximate $g^*$ by
\begin{align}
  g^* \xrightarrow{\omega_q \gg \omega_c} \dfrac{\sqrt{\omega_c\omega_q}}{2}. \label{eq:gstar}
\end{align}
Note this coupling scale has previously been identified as being associated with a qualitative change in the form of the ground state (viewed as a function of $g$)  \cite{2010Nori,2010Choi}.

Returning to the lab frame, the ground state of the approximate Hamiltonian in Eq.~(\ref{eq:BSHam}) is  $e^{-\hat{A}}\hat{S}(-r)\ket{0}\ket{-z}$.
For the regime of interest $\omega_q \gg \omega_q$, $e^{\hat{A}}$ induces a small amount of qubit-cavity entanglement, proportional to the small parameter $g/\omega_q$, but does not affect the magnitude of the squeezing.  This squeezing is the most relevant feature of the weak-coupling ground state, and we will seek to capture it in our variational ansatz.

\section{The ground state}
\label{sec:GSAnsatz}

From our study in the previous section of various limiting cases, we make three key observations about the Rabi-model ground state: {(i)} the cavity state should be squeezed for weak but finite coupling, {(ii)} there is a preference for even-parity reduced cavity states at intermediate couplings, and {(iii)} the large coupling ground state is fully entangled. In this section, using this intuition, we introduce a variational state for the ground state of \cref{eq:RabiHamiltonian} that is a superposition of squeezed coherent states entangled with non-orthogonal qubit-states. Furthermore, we give the Schmidt decomposition of this variational state, and show how our variational state's free parameters can be used to understand the evolution of the Rabi ground state.

\subsection{\label{sec:TheAnsatz} The variational solution for the ground state}

\begin{figure*}[t]
    \centering
    \subfigure{
    \includegraphics[width=0.49\textwidth]{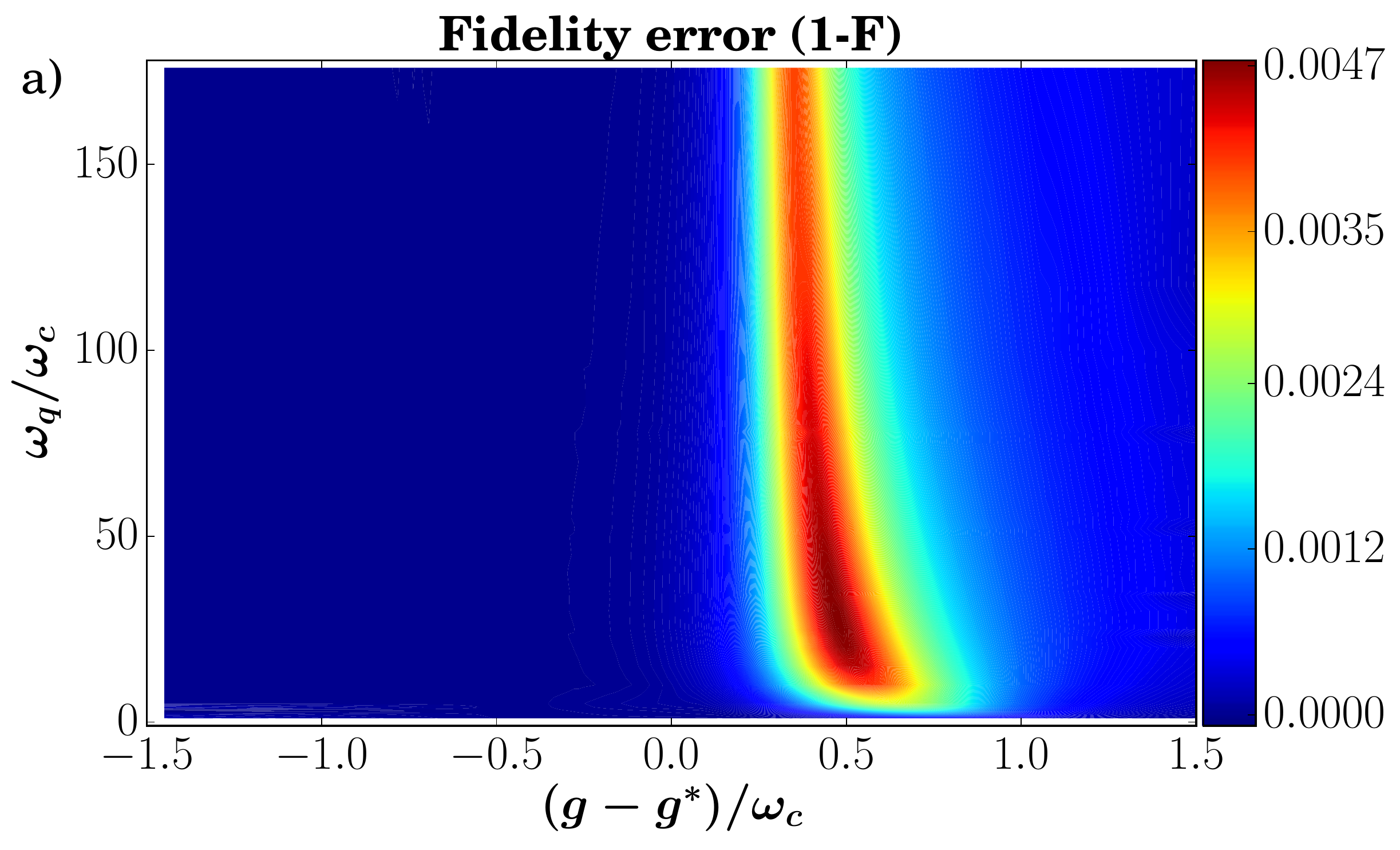}
    \label{fig:GndStateFid}}
    \subfigure{
    \includegraphics[width=0.45\textwidth]{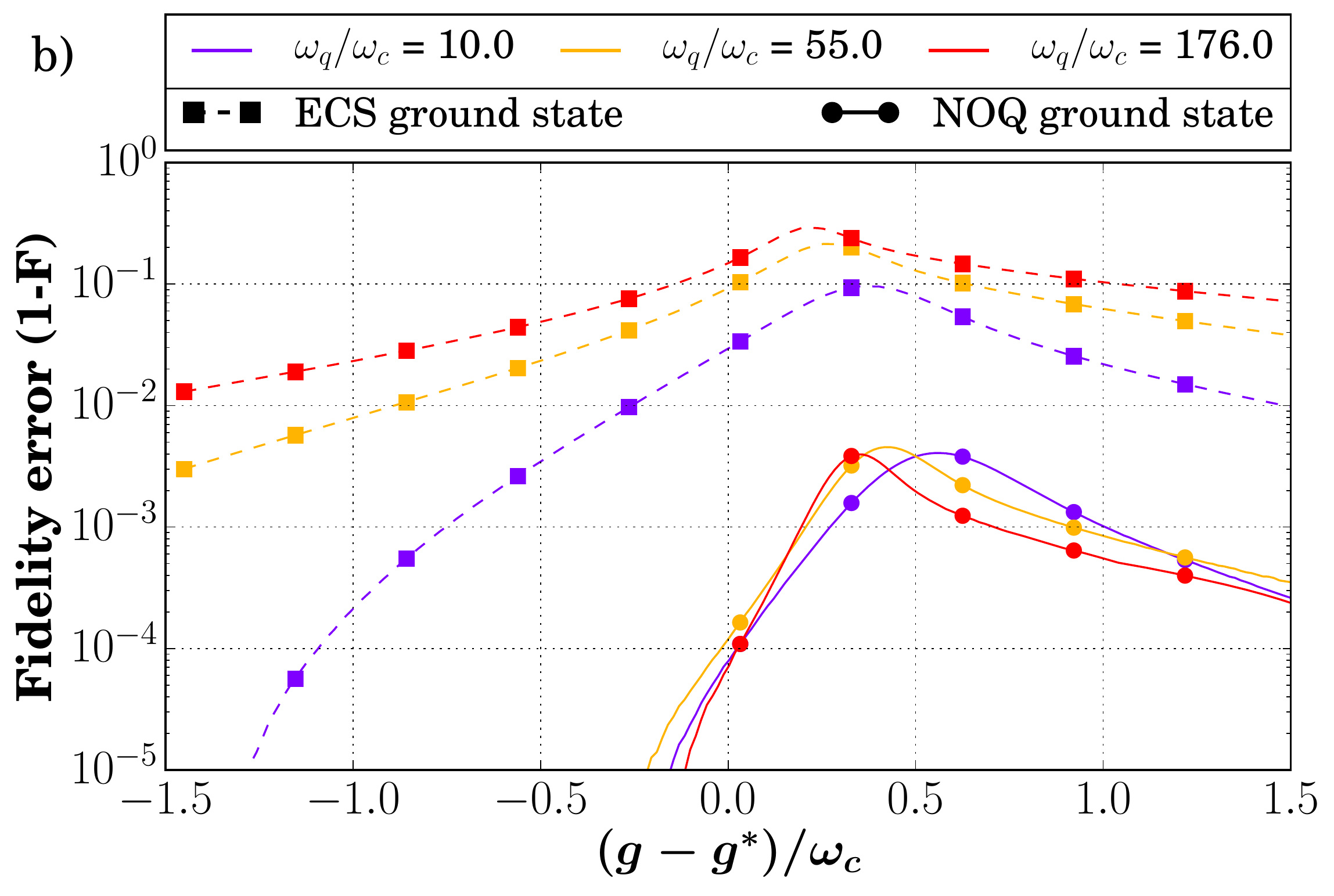}
    \label{fig:GndStateFidC}}
    \caption{{\bf a)} Fidelity error (defined in \cref{eq:Overlap}) for the fidelity-optimized non-orthogonal qubit (NOQ) ground state of \cref{eq:Ansatz}, as a function of coupling strength (defined relative to $g^*$ of \cref{eq:gstar}) and qubit frequency. {\bf b)}  Line-cuts of the fidelity error of the optimal NOQ ground state of \cref{eq:Ansatz} introduced in this paper, and the entangled cat state (ECS) of \cref{eq:ECSg}.}
    \label{fig:GndStateFidA}
\end{figure*}

Our proposed variational state is a simple superposition of two terms, with each being a product of a displaced, squeezed cavity state and a qubit pointer state. Crucially, the two qubit pointer states will in general {\emph not} be orthogonal, due to the preference for even parity at intermediate couplings. We choose the same squeezing parameter for both terms, and pick the displacements to have equal magnitude but opposite sign, such that our variational state is a function of only three free parameters: the displacement $\alpha_c$, the squeezing parameter $r$, and the angle $\phi$ that defines the overlap of the qubit states. This leads to a simple form for our variational state, which we refer to as the non-orthogonal qubit (NOQ) state:
\begin{align}
\ket{\psi_0^{\rm NOQ}[\alpha_c,r,\phi]} = \frac{\ket{\alpha_c,r}\ket{+[\phi]} - \ket{-\alpha_c,r}\ket{-[\phi]}}{\sqrt{2N}} \label{eq:Ansatz}
\end{align}
where
\begin{align}
  &\ket{\alpha_c, r} = \hat{D}(\alpha_c) \hat{S}(-r) \ket{0}, \label{eq:dispsqz} \\
  &\ket{\pm[\phi]} = \cos \left(\frac{\phi}{2}\right) \ket{+z} \mp  \sin \left(\frac{\phi}{2}\right) \ket{-z} \label{eq:SpinState},
\end{align}
and the normalization constant
\begin{equation}
  N = 1 - \exp\left(-2 \alpha_c^2 e^{-2r} \right)\cos \phi.
\end{equation}
The qubit pointer states $\ket{\pm[\phi]}$ satisfy $\braket{\pm[\phi]}{\mp[\phi]} = \cos \phi$; they are formed by symmetrically rotating the state $\ket{+z}$ about the $y$-axis by the angle $\pm\phi$. This form is consistent with the parity-frame variational state of \cref{eq:ParECS}, where the parity frame transformation $\hat{U}_{\Pi}$ is responsible for the symmetric rotation about the $y$-axis.

Our NOQ state is consistent with the known form of the Rabi model ground state in both the weak and strong coupling limits.

For weak coupling, the parameters $\{\alpha_c = 0,\ r=0,\ \phi = \pi\}$ give exactly the fully disentangled ground state reached for $g = 0$, and for small but finite coupling, $g\ll\omega_q$, the NOQ state correctly captures that the ground state is approximately the ground state of \cref{eq:BSHam}, i.e. $\hat{S}(-r)\ket{0}\ket{-z}$. In the large coupling limit, $g\gg\omega_q$, the parameters $\{\alpha_c > 0,\ r=0,\ \phi = \pi/2\}$ give exactly the fully entangled cat state of \cref{eq:ECS}, found for $g\rightarrow\infty$.

The accuracy of the NOQ state in both the strong and weak coupling limits is not surprising, as we explicitly designed the state to be able to do this.  More surprisingly, the NOQ state also captures the true ground state with high-fidelity for \emph{intermediate} values of the coupling, where the standard approximations used for either large or small $g$ fail. As we are interested in faithfully reconstructing the ground state with the NOQ state, we find the optimal parameters, $\{\alpha_c,\ r,\ \phi\}$, by numerically maximizing the fidelity of the NOQ state with the ground state of \cref{eq:RabiHamiltonian} found by numerical diagonalization, using the QuTiP toolbox \cite{qutip} (see \cref{sec:numerics} for further details). The fidelity between the NOQ state and the numerical ground state of the Rabi Hamiltonian is defined in the usual way
\begin{align}
F[\alpha_c,r,\phi] = \abs{\braket{\psi^{\rm NOQ}_0[\alpha_c,r,\phi]}{\psi^{\rm exact}_{0}}} \label{eq:Overlap}
\end{align}
which corresponds to the overlap between the two states. By maximizing the fidelity, our results show just how close the optimally chosen NOQ state is to the exact ground state, even at intermediate values of the coupling.

The results of fidelity optimization are shown in \cref{fig:GndStateFidA}. \cref{fig:GndStateFid} shows a generally high fidelity in the entirety of parameter space examined, with a region of lower fidelity near $g=g^*$. This region of lower fidelity corresponds to the onset of instability in the effective weak-coupling Hamiltonian of \cref{eq:BSHam}, and represents a rough cross-over regime between the weak and strong coupling limits.

We stress that even in this ``low'' fidelity region the NOQ state is still remarkably accurate, and overall, we find a fidelity greater than $99\%$ for all parameters.
\cref{fig:GndStateFidC} shows line-cuts of fidelity versus coupling for selected qubit frequencies.

\begin{figure*}[t]
    \subfigure{
    \includegraphics[width=0.45\textwidth]{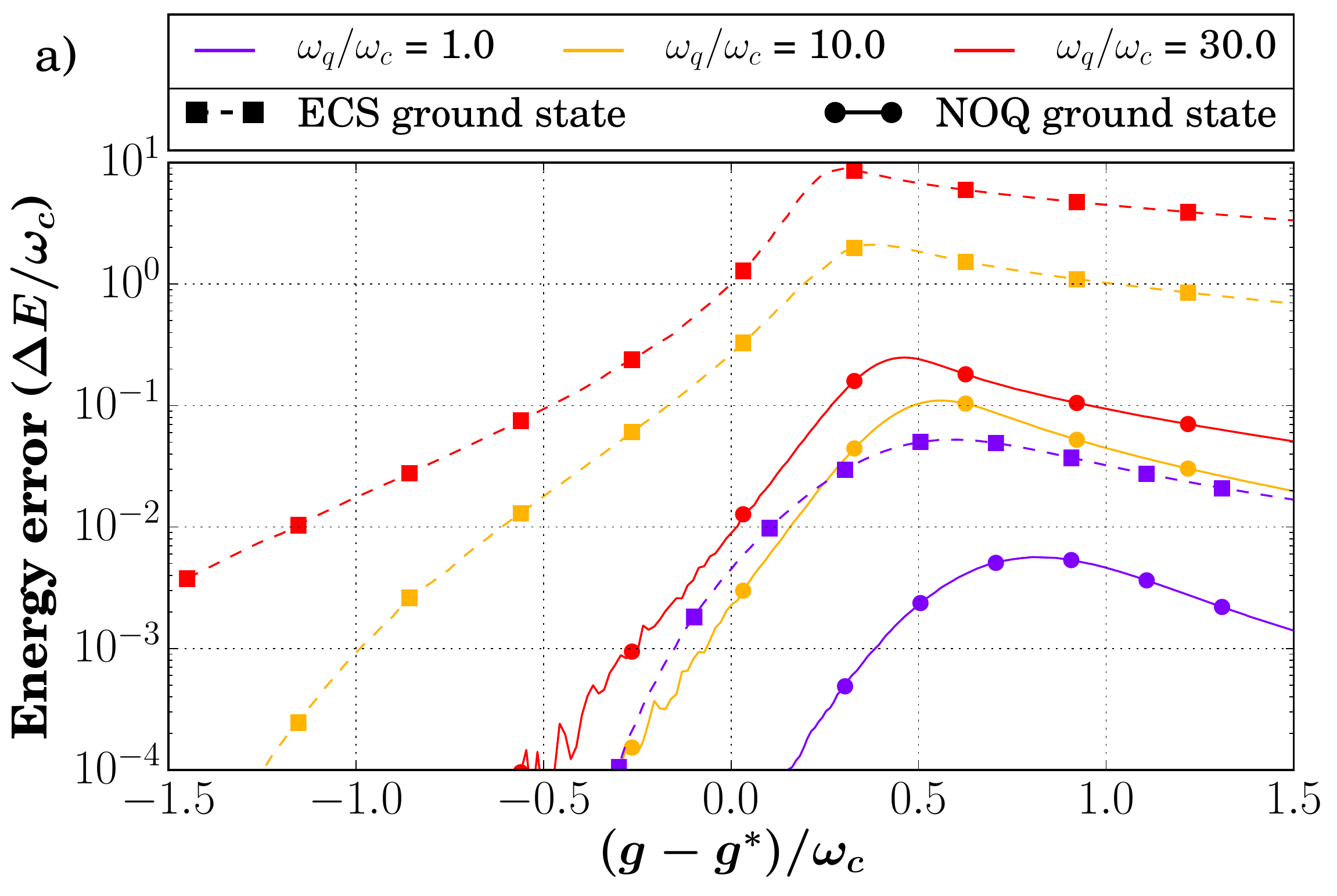}
    \label{fig:GndStateEngC}}
    \subfigure{
    \includegraphics[width=0.45\textwidth]{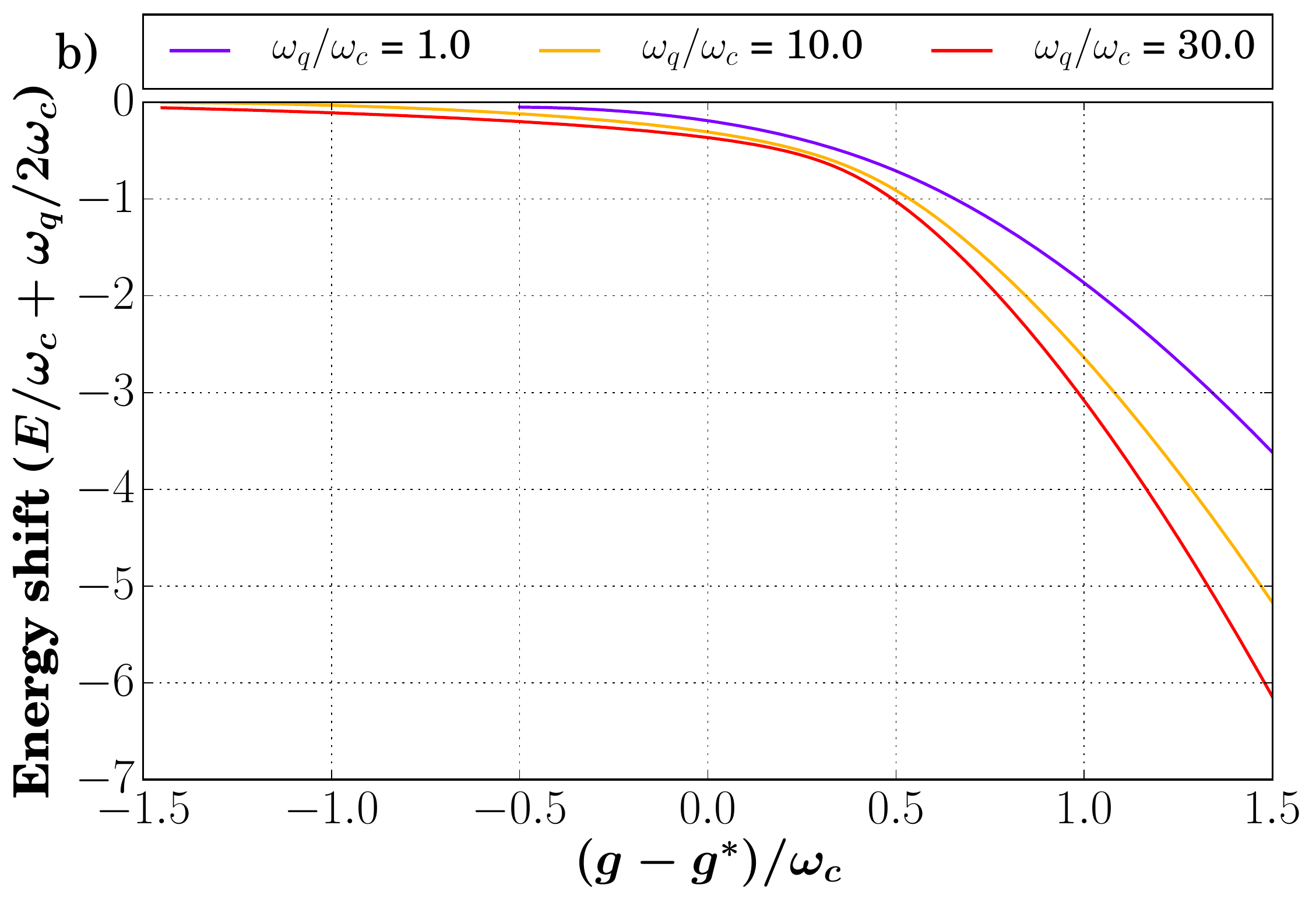}
    \label{fig:GndStateEOffC}}
    \caption{{\bf a)} Energy error (see \cref{eq:DelE}) for the fidelity-optimized non-orthogonal qubit (NOQ) ground state of \cref{eq:Ansatz}, and for the fidelity-optimized entangled-cat state (ECS) of \cref{eq:ECSg}, as functions of coupling strength, for sample qubit frequencies. {\bf b)} The exact energy shift of the ground state energy away from $-\omega_q/2$, as a function of the coupling strength.}
    \label{fig:GndStateEngA}
\end{figure*}

We also compare the energy of our fidelity-optimized NOQ state to that of the true ground state energy by calculating the energy error
\begin{align}
  \Delta E = \bra{\psi^{\rm NOQ}}\hat{H}\ket{\psi^{\rm NOQ}} - \bra{\psi^{\rm exact}}\hat{H}\ket{\psi^{\rm exact}}. \label{eq:DelE}
\end{align}
This error in the ground state energy is shown in \cref{fig:GndStateEngC}. It follows the same trend as the fidelity, with the largest errors
occurring when $g \simeq g^*$. This error should be compared to \cref{fig:GndStateEOffC}, which shows the exact energy shift of the ground state energy away from $-\omega_q/2$, its value at $g=0$. Far from $g=g^*$, the energy error is several orders of magnitude smaller than the exact energy shift.   However, in the region around $g^*$, at its largest the energy error is less than an order of magnitude smaller than the exact energy shift.  This behaviour partially reflects the fact that we have optimized the NOQ state parameters to minimize infidelity with the true ground state, and not average energy.  Minimizing energy (as in a standard variational calculation) would necessarily result in an even lower magnitude of error in the energy estimate.

Indeed, maximizing fidelity with the true ground state to find the optimal parameters of the NOQ state is not practical for future applications of the NOQ state, as it requires the numerically challenging task of diagonalizing the Rabi Hamiltonian. However, we do this here to emphasize clearly how faithfully the NOQ state is able to capture the true ground state.  We describe using the NOQ state in a more standard variational setting (where one minimizes the average energy of the state without any knowledge of the true ground state) in
Appendix \ref{sec:ExpectationValueOfH}.

\begin{figure*}[t]
    \centering
    \subfigure{
    \includegraphics[width=0.45\textwidth]{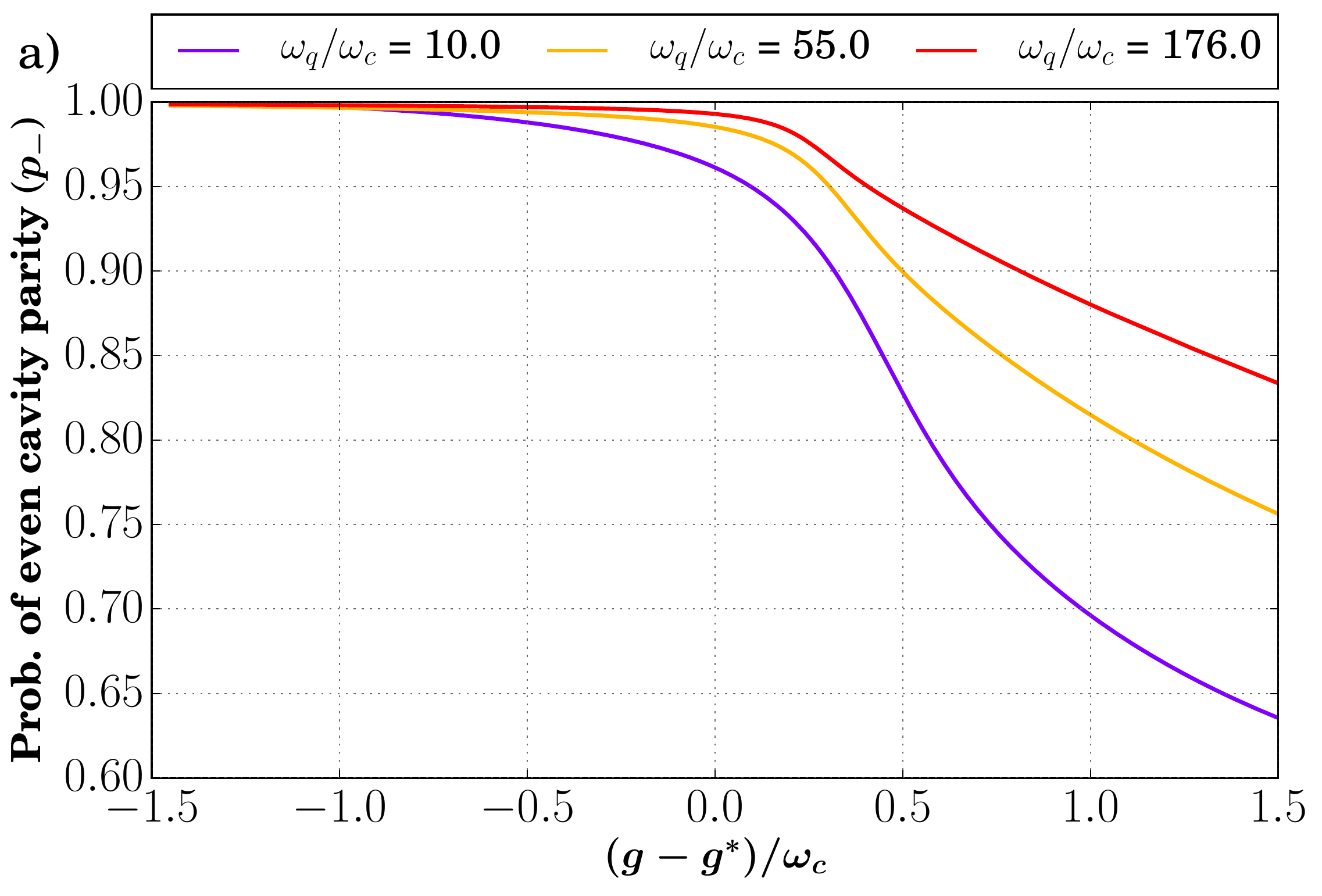}
      \label{fig:GndStatePmz}}
    \subfigure{
    \includegraphics[width=0.45\textwidth]{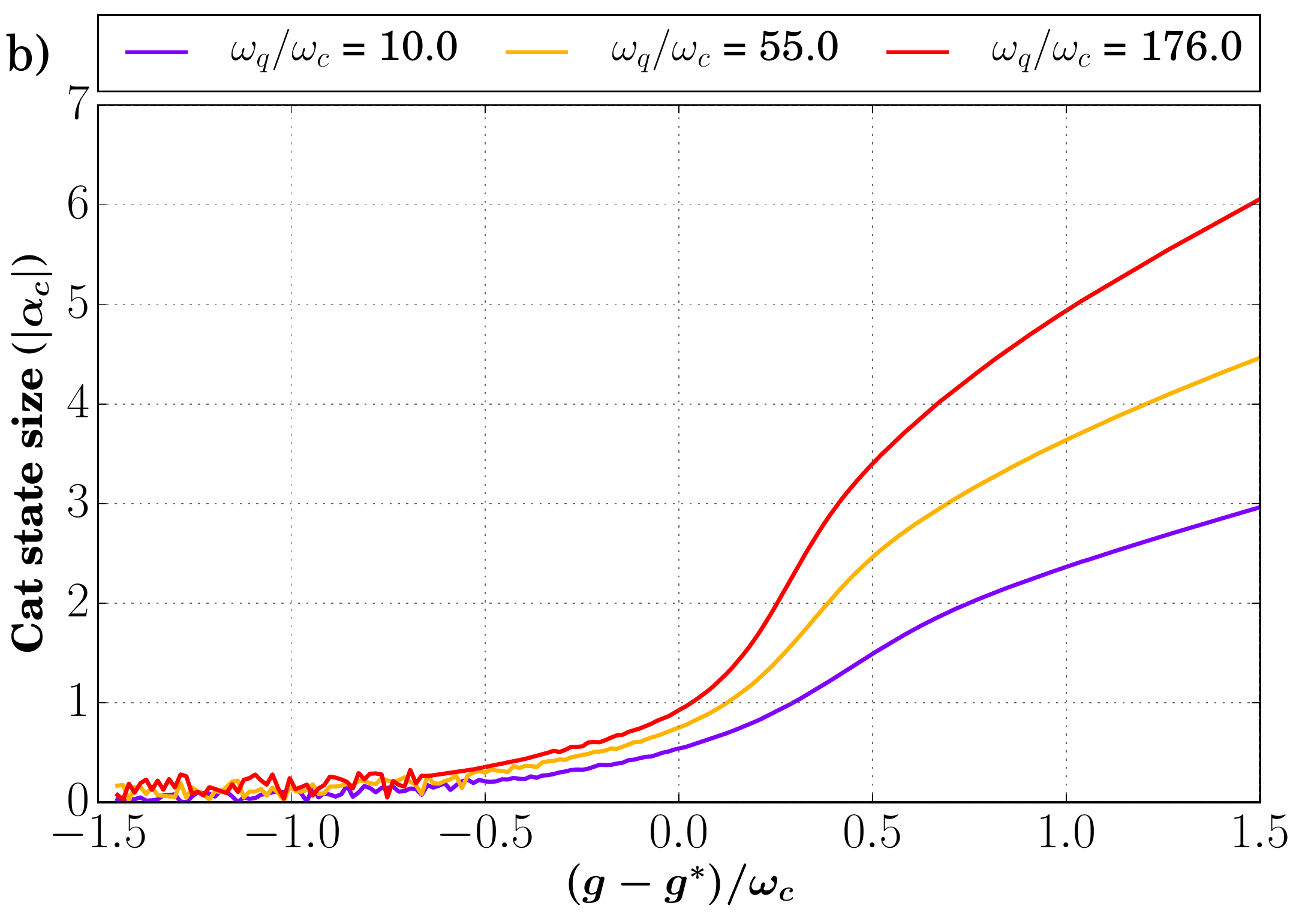}
    \label{fig:GndStateDisp}}
    \subfigure{
    \includegraphics[width=0.45\textwidth]{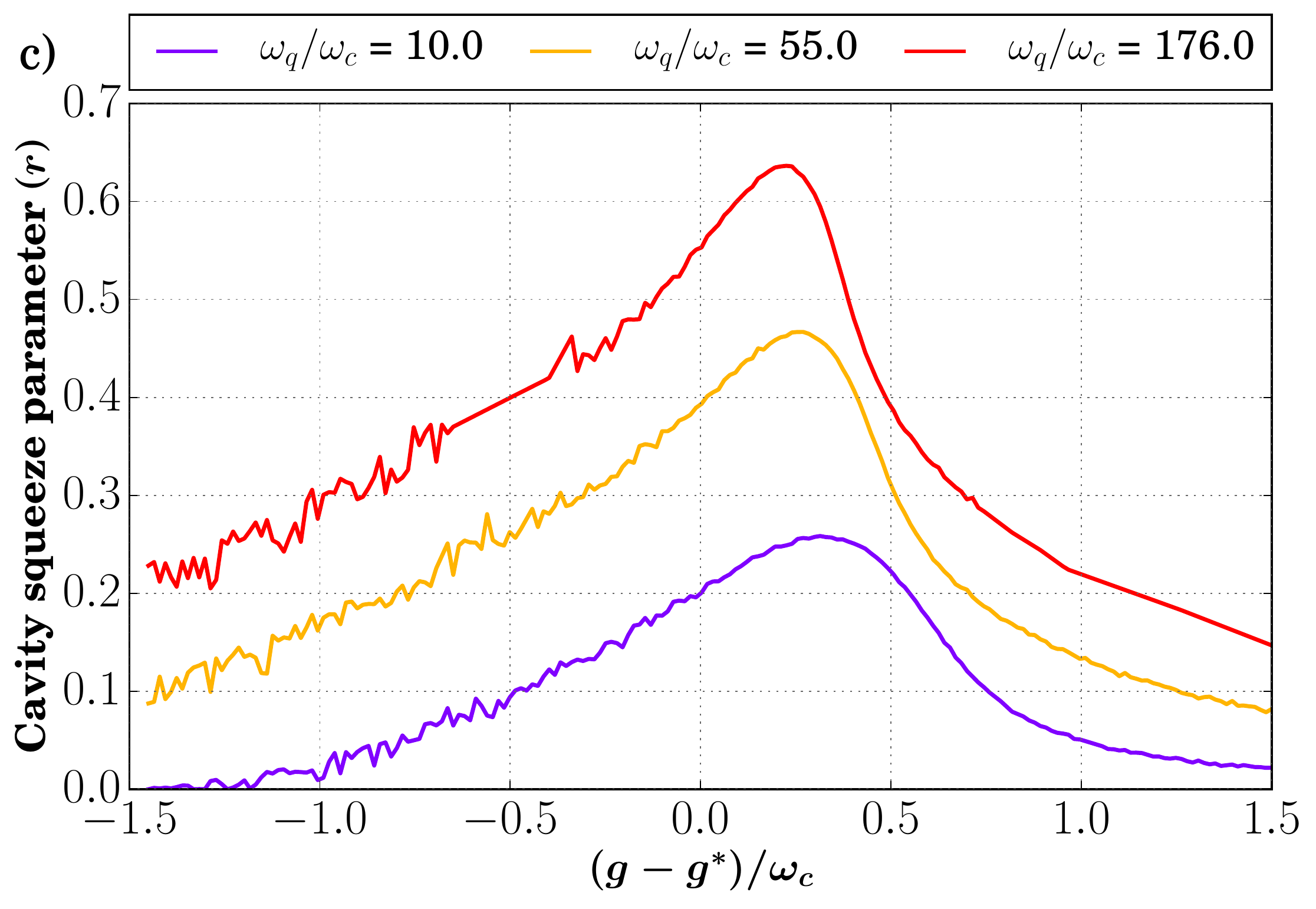}
    \label{fig:GndStateSqz}}
    \caption{Optimal parameters for the non-orthogonal qubit (NOQ) variational state of \cref{eq:Ansatz} found by fidelity maximization, as functions of the coupling strength $g$, for various values of the qubit frequency $\omega_q$. \textbf{a)} shows $p_-$, the probability of the qubit being in its non-interacting ground state $\ket{-z}$ (c.f.~\cref{eq:ProbabilitySpinState}), \textbf{b)} shows the displacement $\alpha_c$, and \textbf{c)} shows the squeezing parameter $r$ of the cavity states. Large values of $\omega_q$ are chosen so that the reduced cavity state is approximately a large Schr\"odinger cat state near the transition regime for $g \simeq g^*$ where $g^* \simeq \sqrt{\omega_c\omega_q}/2$. The noise in \textbf{c)} for small values of $g$ is a result of numerical error.}
    \label{fig:GndStatePar}
\end{figure*}

It is also worth considering other approximate variational solutions for the ground state, and as such we compare the NOQ state to a fully entangled Schr\"odinger cat state (ECS)
\begin{align}
\ket{\psi_-^{\rm ECS}[\alpha]} = \frac{1}{\sqrt{2}}\left(\ket{\alpha}\ket{+x} - \ket{-\alpha}\ket{-x}\right), \label{eq:ECSg}
\end{align}
which is the solution in the limit $g\rightarrow\infty$, and has been previously used to describe the ground state of the Rabi Hamiltonian \cite{Irish:2007aa,Schiro:2013aa}. Such a state has one free parameter, the displacement of the cavity coherent states. Line-cuts of the fidelity error and energy error are shown in \cref{fig:GndStateFidC} and \cref{fig:GndStateEngC} for the fidelity-optimized ECS (i.e.~its overlap with the numerical ground state is maximized). As the plots show, the fidelity-optimized NOQ state is always a better approximation to the exact ground state than the ECS, both in terms of fidelity and the ground state energy error.

\subsection{\label{sec:SchmidtDecomposition}Schmidt decomposition of the variational state}

An alternative description of our NOQ state is obtained if we write it in Schmidt form, which is a superposition with components built from orthonormal cavity and qubit states. The qubit states are simply the eigenstates of $\hat{\sigma}_z$, while the cavity states are squeezed cat states, and the interesting behavior is now contained in the Schmidt coefficients describing the weighting of the superposition. In Schmidt form, the NOQ state of \cref{eq:Ansatz} is
\begin{align}
\nonumber\ket{\psi^{\rm NOQ}_0[\alpha_c,r,\phi]} &= \sqrt{p_-} \ket{\Phi^S_+[\alpha_c,r]}\ket{-z} \\&+ \sqrt{1-p_-} \ket{\Phi^S_-[\alpha_c,r]} \ket{+z}, \label{eq:SchmidtDecomposition}
\end{align}
where
\begin{equation}
\ket{\Phi^S_{\pm}[\alpha_c,r]} = \frac{1}{\sqrt{2 \mathcal{N}_\pm}} \left(\ket{\alpha_c,r} \pm \ket{-\alpha_c,r}\right), \label{eq:sqzCat}
\end{equation}
are squeezed cat states with opposite parity, with normalizations $\mathcal{N}_\pm = 1 \pm \exp\left(-2\alpha_c^2 e^{-2r}\right)$. The Schmidt coefficients are defined by the parameter $p_-$, the probability of the qubit being in the state $\ket{-z}$:
\begin{equation}
p_-  = \frac{1}{2} - \frac{\exp\left(-2\alpha_c^2e^{-2r}\right) - \cos \phi}{2N^2}, \label{eq:ProbabilitySpinState}
\end{equation}
and satisfies $p_- \geq 0.5$ since $\cos \phi \leq 0$.

The Schmidt decomposition of the NOQ state explicitly shows that this state always has an even total parity.  \Cref{eq:SchmidtDecomposition} is in fact the variational state of \cref{eq:ParECS} introduced as a guess for the ground state in the large coupling limit, with additional squeezing added to improve the fidelity with the exact state for weak coupling.

The advantage of writing the NOQ state in Schmidt form is that the evolution of the ground state is very clearly described by the three parameters $\{p_-,\ \alpha_c,\ r\}$. In \cref{fig:GndStatePar} we plot as functions of $g$ the values of these parameters for the NOQ state which gives maximum fidelity with the true ground state.  Examining the limiting cases, we see that for weak coupling, where the cavity ground state is approximately squeezed vacuum, we find $\{p_- \simeq 1,\ \alpha_c \simeq 0,\ r>0\}$.  As the coupling $g$ is increased further, the
displacement $\alpha_c$ increases, with a more rapid increase beginning around $g = g^*$. This is explained by the onset of instability in the second order perturbative Hamiltonian of \cref{eq:BSHam} (which implies that the qubit-cavity interaction can no longer be treated perturbatively).
As $g$ becomes extremely large, we have $\{p_- \rightarrow 1/2,\ \alpha_c \rightarrow g/\omega_c,\ r = 0\}$ as $g\rightarrow \infty$, which tends to the expected strong coupling ground state (the entangled cat state of \cref{eq:ECSg}).  It is interesting to note that the squeezing correlations in the ground state (described by $r$) are strongest near the crossover value of the coupling $g = g^*$; while they are never large, keeping these correlations is crucial in order to obtain a high fidelity with the exact ground state.

The Schmidt form of our NOQ state and \cref{fig:GndStatePar} also provide a hint as to why the ground state corresponds to a near-pure cavity cat state when $\omega_q \gg \omega_c$ and $g \sim g^*$.  In this cross-over regime, the coherent displacement $\alpha_c$ can already have a large magnitude, while the probability $p_{-}$ of having the qubit in its (non-interacting) ground state $\ket{-z}$ remains close to one.  As the ground state must have an even total excitation parity (c.f.~discussion in Sec.~\ref{sec:limits}), it follows that the cavity is close to being in a pure (squeezed) cat state (as described by the first term in \cref{eq:SchmidtDecomposition}).

\section{Schr\"{o}dinger Cat State Formation in the Quantum Rabi Model}
\label{sec:CatStates}

\cref{fig:WignerPlots} and \cref{fig:GndStatePar} demonstrate a surprising effect:  there exist parameter regimes where the Rabi-model ground state corresponds to the cavity being in a relatively pure, large-displacement (and slightly squeezed) cat state.  This occurs when $\omega_q \gg \omega_c$, and when $g$ is near the cross-over scale $g^*$.

The qualitative change in the ground state wavefunction near $g \sim g^*$ can be associated with the onset of instability of the weak-coupling effective squeezing Hamiltonian  of \cref{eq:BSHam} (see Sec.~\ref{sec:limits}).  For $g \simeq g^*$, higher-order qubit-induced cavity  nonlinearities are required to maintain stability; such nonlinearities would naturally lead to non-trivial cavity states.
\cref{fig:WignerPlots} and \cref{fig:GndStatePar} demonstrate that while this effective nonlinearity arises from interaction with the qubit, it does \emph{not} require strong qubit-cavity entanglement. Only a small amount of qubit-cavity entanglement/hybridization is required to generate a relatively large effective nonlinearity in the cavity.  The result is non-trivial cavity states that remain relatively pure.

A complementary understanding of these pure cavity cat states can be obtained by using the Schmidt form of our NOQ state
(c.f. \cref{eq:SchmidtDecomposition}.
 In the regime $g \simeq g^*$, which corresponds to ultrastrong coupling, the weights of the states $\ket{\Phi^S_{\pm}}$ in \cref{eq:SchmidtDecomposition} are imbalanced (i.e. $p_- > 1-p_-$), and the cavity state is thus approximately the even parity cat state $\ket{\Phi^S_{+}}$, with $\alpha_c >0$.  This imbalance is due to the large value of $\omega_q$, which gives an energetic preference for the qubit to be in the $\ket{-z}$ state.  As the displacement $\alpha_c$ grows with increasing $g$, large-displacement cat states are possible for larger $\omega_q$, as this pushes the beginning of the transition regime defined by $g^* \approx \sqrt{\omega_c\omega_q}/2$ to larger values of $g$.

\begin{figure}[h!]
  \centering
  \includegraphics[width=0.45\textwidth]{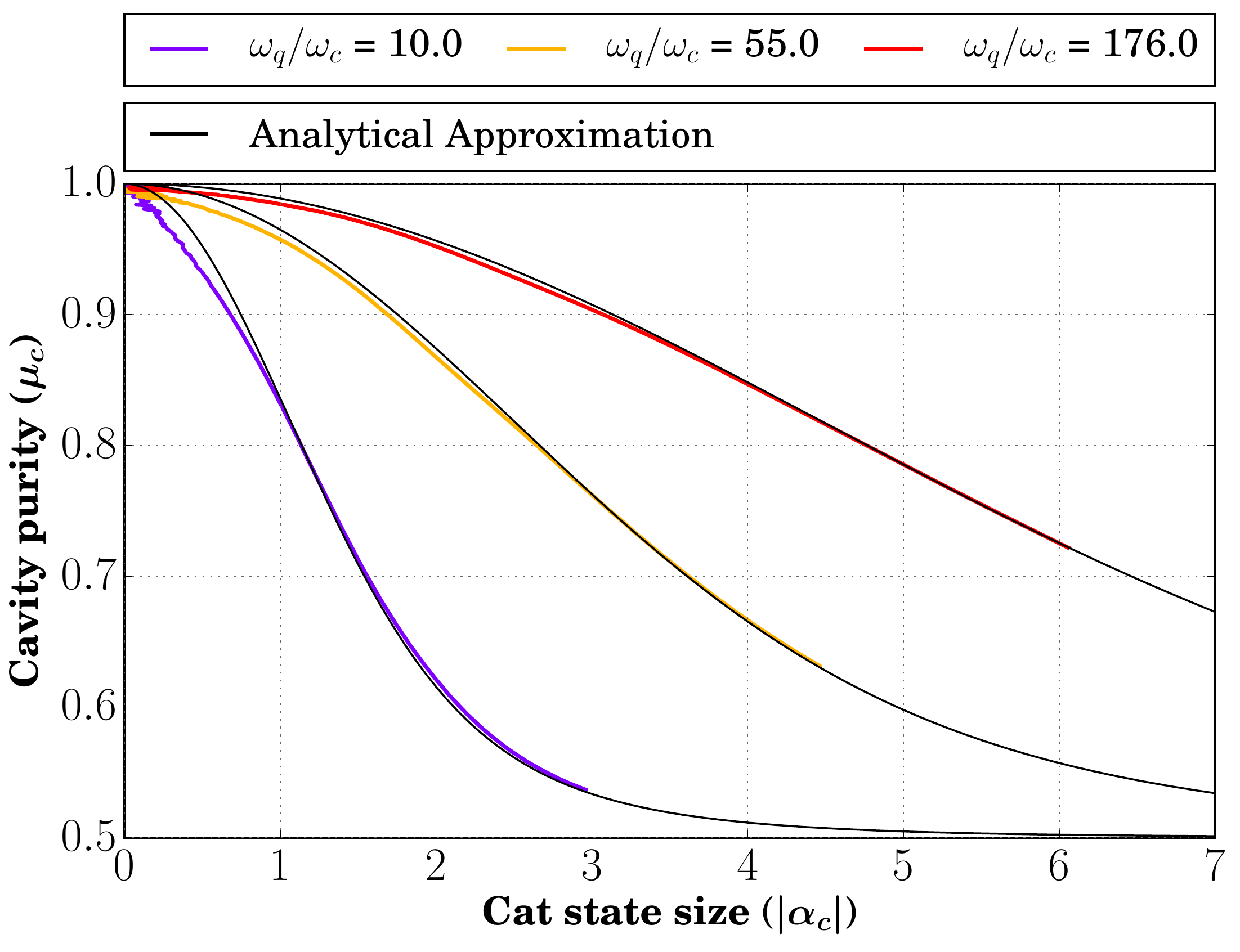}
  \caption{\textbf{a)} For three sample values of $\omega_q$, we increase the coupling strength $g$, and for each value of $g$ find the fidelity-optimized non-orthogonal qubit (NOQ) state of \cref{eq:Ansatz}. For the NOQ state at each value of $g$, this plot shows the purity of the reduced cavity state, $\mu_c$ of \cref{eq:purity}, and the size of the cavity cat state displacement, $\abs{\alpha_c}$ of \cref{eq:sqzCat}. For large values of $\omega_q$, the reduced cavity state can have both high purity and large displacement. The numerical results (colored lines) are compared to the analytical expressions of \cref{eq:approx_purity} and \cref{eq:approx_alpha} (black lines).}
  \label{fig:PurDisp}
\end{figure}

To measure how close the reduced cavity state is to a pure cat state, we calculate the purity of the reduced cavity state of the NOQ state, given by
\begin{equation}
\mu_c = (1-p_-)^2 + p_-^2.  \label{eq:purity}
\end{equation}
As the impurity of the cavity is solely due to qubit-cavity entanglement, the cavity purity is also a measure of qubit-cavity entanglement. \cref{fig:PurDisp} shows the purity for a given displacement $\alpha_c$.  One sees clearly that for large
 $\omega_q / \omega_c$, large-magnitude displacements can be achieved before the purity drops appreciably from unity.

Treating the NOQ state as a standard variational ansatz, and determining its optimal parameters by minimizing the ground state energy (see Appendix \ref{sec:ExpectationValueOfH} for more details), it is possible to determine approximate analytical expressions for the purity and $\alpha_c$ after the transition point $g=g^*$. By minimizing the expectation value of the Rabi Hamiltonian with the NOQ state of \cref{eq:SchmidtDecomposition}, we find the approximate solutions
\begin{align}
&\mu_c \simeq \frac{1}{2} \left( 1 + \left(\frac{\sqrt{\omega_c\omega_q}}{2g}\right)^4\right), \label{eq:approx_purity} \\
&\alpha_c \simeq  \frac{g}{\omega_c} \sqrt{1 - \left(\frac{\sqrt{\omega_c\omega_q}}{2g}\right)^4}, \label{eq:approx_alpha}
\end{align}
and from these we can write the purity as a function of the displacement
\begin{align}
\mu_c \simeq \left(1 + \frac{1}{\sqrt{1+\left(\frac{\omega_q/\omega_c}{2\alpha_c^2}\right)^2}}\right)^{-1}. \label{eq:purdisp}
\end{align}
This equation shows that in general, increasing $\alpha_c$ decreases the purity, but for fixed $\alpha_c$, increasing $\omega_q$ will increase the purity. Therefore, large displacements and high purity are possible provided $\omega_q$ is large enough. Using this expression, we can say that, as an example, for $90\%$ purity one needs $\alpha_c \simeq 0.24\sqrt{\omega_q/\omega_c}$.  If we take $\omega_q/\omega_c = 176$, this implies that the ``size" of our cat is approximately  $\abs{\alpha_c} \simeq 3.14$.  In contrast, if we reduce $\omega_q$ so that $\omega_q/\omega_c = 10$, the size of the cat drops to  $\abs{\alpha_c}<0.75$.  This quantitively shows that high purity is compatible with large ``cat sizes" $\abs{\alpha_c}$, provided that $\omega_q / \omega_c$ is large enough.  An example Wigner function of such a high-purity, relatively large cat state is given in \cref{fig:scheme}(b). This cat state is also slightly squeezed, with a squeeze parameter $r \simeq 0.345$.

\section{Variational solution for the first excited state}
\label{sec:HigherLevels}

It is also possible to extend our NOQ variational ground state ansatz to accurately describe the first excited state of the Rabi Hamiltonian (c.f. \cref{eq:RabiHamiltonian}). To that end, we first consider the known limits of the first excited state. For a large $\omega_q$, the qubit is effectively frozen in the $\ket{-z}$ state for low energy eigenstates, as the energy cost for the qubit to be excited is too great. Thus, in the weak coupling regime, $g\ll g^*$, the first excited state of the Rabi Hamiltonian is approximately described by the state $\hat{S}(r)\ket{1}\ket{-z}$, where the cavity state is a squeezed one photon Fock state, with squeeze parameter $r$ of \cref{eq:sqzr}, as was discussed for the ground state in section \ref{sec:limits}.

In the large coupling regime $g \gg g^*$, as described in section \ref{sec:limits}, the large coupling strength results in an effective qubit-state dependent cavity displacement, and the two lowest energy eigenstates are described by the degenerate qubit-cavity entangled cat states \cite{Irish:2007aa}, shown in \cref{eq:ECS}. The first excited state is
\begin{align}
\ket{\psi_{+}^{\rm ECS}[\alpha]} = \frac{1}{\sqrt{2}}\left(\ket{\alpha}\ket{+x}+\ket{-\alpha}\ket{-x}\right), \label{eq:ECSe}
\end{align}
which just like the ground state solution is a fully-entangled cat state (but now with odd parity).

As was done for the variational ground state, we introduce a variational first excited state that captures both limits by incorporating both squeezing and displacement as variational parameters. Our variational first excited state takes the form
\begin{align}
\ket{\psi^{\rm NOQ}_{1}[\alpha_c,r,\phi]} = \frac{\ket{\alpha_c,r}\ket{+[\phi]} + \ket{-\alpha_c,r}\ket{-[\phi]}}{\sqrt{2N_1}}, \label{eq:1E}
\end{align}
with $\ket{\alpha_c,r}$ the displaced squeezed states of \cref{eq:dispsqz}, and $\ket{\pm[\phi]}$ the nonorthogonal qubit states of \cref{eq:SpinState}, with normalization now given by
\begin{equation}
N_1 = 1 + \exp\left(-2\alpha_c^2e^{-2r}\right) \cos\phi.
\end{equation}
The NOQ first excited state has a similar form to the NOQ ground state of \cref{eq:Ansatz}, with the only difference being the relative phase between the superposition components. However, fidelity optimization will not not necessarily obtain the same values of the free parameters $\{\alpha_c,r,\phi\}$ for the ground state as the first excited state, and as such the NOQ ground state and NOQ first excited state are not orthogonal by design (as would be the case in a standard variational approach). They can be made orthogonal by introducing this as a constraint to the optimization algorithm, and while this was not done in the present work, it can easily be implemented for future work that involves both variational eigenstates.

\begin{figure}[t]
    \centering
    \includegraphics[width=0.45\textwidth]{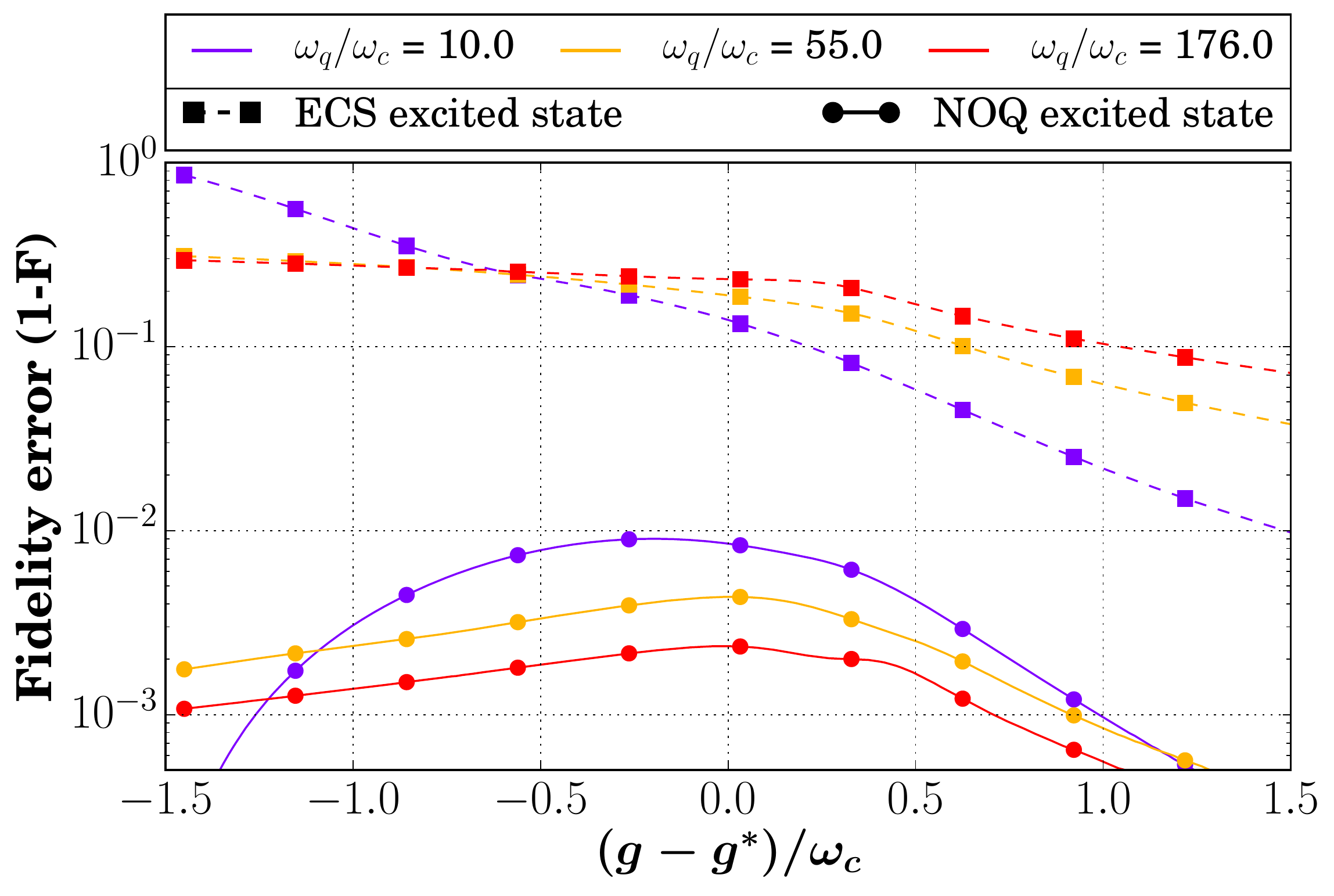}
    \caption{Line-cuts of the fidelity error for the fidelity-optimized non-orthogonal qubit (NOQ) first excited state of \cref{eq:1E}, and the fidelity-optimized entangled cat state (ECS) solution of \cref{eq:ECSe}. The NOQ first excited state is a better approximation than the ECS solution for all values of $g$ and $\omega_q$ considered.}
    \label{fig:1stExcitedState}
\end{figure}

As is shown in \cref{fig:1stExcitedState}, the fidelity-optimized NOQ first excited state has high fidelity with the exact numerical eigenstate, and is a better approximation than the fidelity-optimized ECS solution in all parameter regimes, even for large coupling.
Unfortunately, variational states similar to those we have proposed for the ground and first excited state do not work for higher eigenstates. In the large coupling regime they are no better than the ECS solutions, while in the weak coupling regime they perform very poorly. Nevertheless, for many applications only the ground and first excited state are needed \cite{Schiro:2012aa,Schiro:2013aa}, and the NOQ states offer an accurate and simple description that can be applied in these situations.

\section{Conclusion}

We have introduced a new variational ground state for the quantum Rabi model that has high fidelity with the exact ground state. Our variational solution is motivated by the properties of the exact ground state in analytically tractable parameter regimes, and is conceptually simple, with only three free parameters.  Further, it only makes use of Gaussian cavity states. As the variational state is conceptually simple, it enables a better understanding of the nature and evolution of the exact ground state as a function of system parameters.

We have used our variational state to describe the evolution of the  ground state with increasing qubit-cavity coupling as a transition from a ``dark" state with zero average photon number, to a ``bright" state with a large average photon number. For intermediate coupling strengths, corresponding to ultra-strong coupling, the ground state yields a reduced cavity state that is approximately a pure Schr\"odinger cat state. Our variational ground state gives a clear intuitive picture for why such states can exist, highlighting that the qubit-induced effective cavity nonlinearity is much stronger than the qubit-cavity entanglement at ultrastrong coupling. In addition, our variational state quantifies the parameter regimes required for cat states, and we find that for a large qubit frequency compared to the cavity frequency, high-purity cat states with large amplitude can be found in the cavity.

In future work, it would be interesting to attempt to use our NOQ variational ground state as a starting point for treating more complicated models, such as Rabi-lattice models (i.e.~a lattice where each site is a Rabi model, and there is tunneling between cavities), or multi-mode ultra-strong coupling systems, such as have been recently demonstrated experimentally \cite{Forn-Diaz:2017aa}.

\section*{Acknowledgement}

This work was supported by NSERC.


\begin{figure*}[t!]
    \centering
    \subfigure{
    \includegraphics[width=0.45\textwidth]{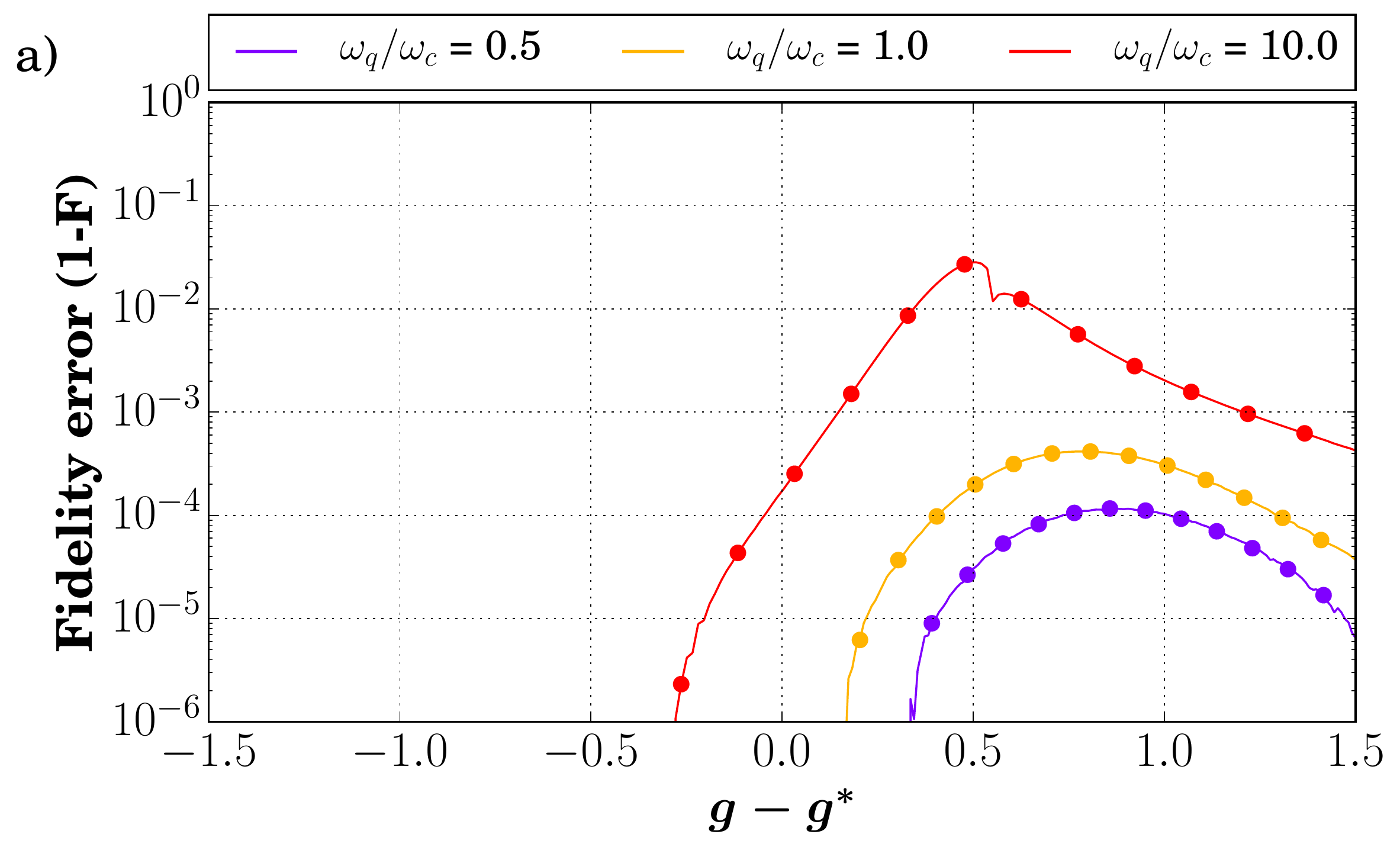}
      \label{fig:EMGSFE}}
    \subfigure{
    \includegraphics[width=0.45\textwidth]{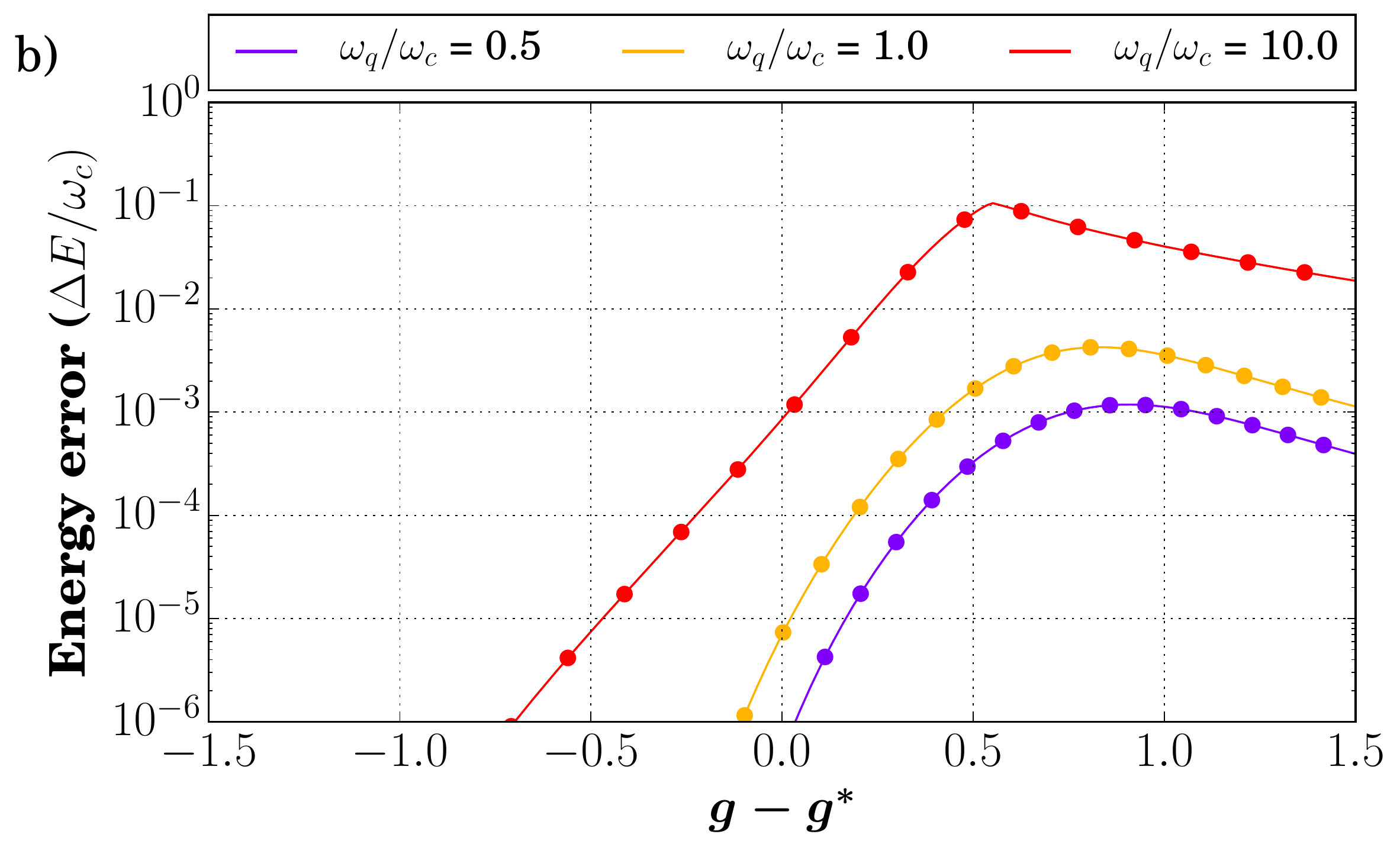}
    \label{fig:EMGSEE}}
    \caption{{\bf a)} Fidelity error (see \cref{eq:Overlap}) and {\bf b)} energy error (see \cref{eq:DelE}) for the non-orthogonal qubit (NOQ) ground state of \cref{eq:Ansatz} as a function of coupling strength (defined relative to $g^*$ of \cref{eq:gstar}). The optimal parameters for the NOQ state were now found by minimization of its expectation value with the Rabi Hamiltonian (see \cref{eq:ExpValParity}).}
    \label{fig:GndStateEMin}
\end{figure*}

\appendix

\section{Numerical Fidelity Maximization}
\label{sec:numerics}

The numerical maximization of the fidelity defined in \cref{eq:Overlap} was done using the differential evolution (DE) algorithm. DE is a stochastic population based method that iteratively improves candidate solutions to globally optimize a given problem \cite{Storn1997}. The advantages of DE over gradient methods, such as gradient descent, are threefold: (i) it is less sensitive to local extrema of the multivariate function, (ii) it converges faster to the global minimum, and (iii) it doesn't require an initial guess. Additionally DE can be used for functions that are not continuous, unlike gradient descent.

For any numerical treatment, the cavity Hilbert space must be truncated at some finite value. The truncation of the cavity Hilbert space, i.e. the largest Fock state considered in our simulations, is determined from the entangled cat state solution of \cref{eq:ECSg}, which is the ground state when $\omega_q=0$, as this state requires the largest cavity Hilbert space to describe. To see that this is the case, we note that the largest displacement in the cavity is for $g/\omega_q \rightarrow \infty$, i.e. the ECS state, and that the squeezing in the cavity at weak coupling is expected to remain small, and crucially is only present when there is not any appreciable displacement.

For $\omega_q=0$ the reduced cavity state is a mixture of coherent states, which makes analytic calculations simple. The matrix element for the component $\ket{n}\bra{n}$ is given by
\begin{equation}
p(n) = \frac{e^{-\abs{\alpha}^2}\abs{\alpha}^{2n}}{n!}, \label{eq:pn}
\end{equation}
and the average photon number is
\begin{align}
  \abs{\alpha}^2 = \expectationvalue{\hat{a}^{\dagger}\hat{a}}= \left(\frac{g}{\omega_c}\right)^2.
\end{align}

For large $\abs{\alpha}^2$, we choose the dimension of the cavity Hilbert space to be the closest integer to $5\abs{\alpha}^2$, which keeps all Fock states with significant matrix elements. For small $\abs{\alpha}^2$, we set a lower bound to the cavity Hilbert space dimension of 50, which captures the comparatively more persistent tail of the distribution of \cref{eq:pn} for small $\abs{\alpha}^2$. These choices are sufficient to accurately describe the numerical ground state, and we have confirmed this fact by varying the Hilbert space dimension in our simulations to test for convergence.

\section{ground state Energy Minimization \label{sec:ExpectationValueOfH}}

In order to perform a true variational approach, and minimize the expectation value of the Rabi Hamiltonian with the NOQ state, it is easier to work in the parity frame defined in section \ref{sec:limits}. The NOQ state of \cref{eq:SchmidtDecomposition} in the parity frame reads
\begin{align}
\nonumber&\ket{\psi^{\Pi \ \rm NOQ}_0[\alpha_c,r,\phi]} \\
&= \left(\sqrt{p_-} \ket{\Phi^S_+[\alpha_c,r]} - \sqrt{1-p_-} \ket{\Phi^S_-[\alpha_c,r]}\right) \ket{-z} \label{eq:NOQStateParity}
\end{align}
where $\ket{\Phi^S_\pm[\alpha_c,r]}$ are defined in \cref{eq:sqzCat}. Its expectation value with the parity frame Hamiltonian of \cref{eq:effectiveH} is
\begin{align}
\expectationvalue{\hat{H}_{\hat{\Pi}}} = \omega_c \expectationvalue{\hat{a}^{\dagger}\hat{a}} + g \expectationvalue{\hat{a}+\hat{a}^{\dagger}} - \omega_q \left(p_--\frac{1}{2}\right),
\label{eq:ExpValParity}
\end{align}
where, the average photon number is
\begin{align}
&\expectationvalue{\hat{a}^{\dagger}\hat{a}} =  \sinh^2 r \\ \nonumber&+  \alpha_c^2 e^{-2r} \cosh 2r
 \left( \frac{\mathcal{N}_-}{\mathcal{N}_+} p_- + (1-p_-)\frac{\mathcal{N}_+}{\mathcal{N}_-} + \tanh 2r \right),
\end{align}
and the net displacement is
\begin{equation}
\expectationvalue{\hat{a}+\hat{a}^{\dagger}} = - 4 \alpha_c \sqrt{\frac{p_-(1-p_-)}{\mathcal{N}_+\mathcal{N}_-}}.
\end{equation}
As in a standard variational calculation, minimization of \cref{eq:ExpValParity} is done numerically with respect to $\alpha_c$, $r$, and $p_-$, and this was also done by differential evolution.

\cref{eq:approx_purity} and \cref{eq:approx_alpha} were derived by this energy minimization approach in the limit $g > g^*$, where approximate analytical solutions are possible. For $g > g^*$, where the cavity displacement $\alpha_c$ is large and the squeezing $r\sim 0$ becomes negligible, we have $\exp(-2\alpha_c^2e^{-2r}) \simeq 0$, such that
\begin{align}
\expectationvalue{\hat{H}_{\hat{\Pi}}} \simeq \omega_c \alpha_c^2- 4 g \alpha_c\sqrt{p_-(1-p_-)} - \omega_q p_- + \rm cst. \label{eq:LGH}
\end{align}
Minimizing \cref{eq:LGH} with respect to $\alpha_c$ yields
\begin{align}
\alpha_c \simeq \frac{2 g}{\omega_c} \sqrt{p_-(1-p_-)}, \label{eq:LGAlpha}
\end{align}
while minimizing with respect to $p_-$ results in
\begin{align}
\nonumber p_- &\simeq  \frac{1}{2} \left(1 - \frac{2\alpha_c^2 \omega_c}{\omega_q} + \sqrt{1+\left(\frac{2\alpha_c^2\omega_c}{\omega_q}\right)^2}{}  \right) \\
&\simeq  \frac{1}{2}\left(1+\frac{\omega_c\omega_q}{4g^2}\right), \label{eq:LGP}
\end{align}
where we have also used \cref{eq:LGAlpha}. Together, these equations allow us to derive the expression in \cref{eq:approx_alpha} for the cavity displacement
\begin{equation}
\alpha_c \simeq \frac{g}{\omega_c} \sqrt{1-\left(\frac{\omega_c\omega_q}{4g^2}\right)^2},
\end{equation}
and the expression in \cref{eq:approx_purity} for the purity
\begin{align}
\mu_c = p_-^2 + (1-p_-)^2
\simeq & \frac{1}{2}\left(1+\left(\frac{\omega_c\omega_q}{4g^2}\right)^2\right).
\end{align}

\section{Relation to other variational solutions}
\label{sec:Comp}

In Ref.~\cite{2010Choi}, Hwang and Choi introduce a variational ground state that they refer to as the double squeezed state (DSS), which in the lab frame takes the form
\begin{align}
  &\ket{\psi_0^{\rm DSS}\left[\alpha_1,\ \alpha_2,\ r_1,\ r_2,\ t\right]}=  \label{eq:ChoiS} \\
  \nonumber&\frac{1-t}{2N_D}\hat{S}(r_1)\left[\left(\ket{\alpha_1} + \ket{-\alpha_1}\right)\ket{-z} + \left(\ket{\alpha_1} - \ket{-\alpha_1}\right)\ket{+z}\right] \\
  &\nonumber+ \frac{t}{2N_D}\hat{S}(r_2)\left[\left(\ket{\alpha_2} + \ket{-\alpha_2}\right)\ket{-z} + \left(\ket{\alpha_2} - \ket{-\alpha_2}\right)\ket{+z}\right],
\end{align}
where $N_D$ is a normalization factor that depends on all free parameters. In general, this state has five free parameters, $\{\alpha_1,\ \alpha_2,\ r_1,\ r_2,\ t\}$, but for $\alpha_2 = -\alpha_1$ and $r_1 = r_2$, it reduces to
\begin{align}
  &\ket{\psi_0^{\rm DSS}\left[\alpha_1,\ -\alpha_1,\ r_1,\ r_1,\ t\right]} = \label{eq:ChoiSr} \\
  \nonumber&\hat{S}(r_1)\frac{\left(\ket{\alpha_1} + \ket{-\alpha_1}\right)\ket{-z} + (1-2t)\left(\ket{\alpha_1} - \ket{-\alpha_1}\right)\ket{+z}}{2N_D}.
\end{align}
For this specific case, the DSS is the same as the NOQ state, as can be seen when one compares \cref{eq:ChoiSr} to \cref{eq:SchmidtDecomposition}. With the identification $\sqrt{1- p_-} = (1-2t)/2N_D$ it is clear that the $t$ parameter in the DSS plays a similar role to the $p_-$ (or $\phi$) parameter in the NOQ state. In light of this fact, one can consider our NOQ state to be a special case of the double squeezed state proposed in Ref.~\cite{2010Choi}, as it is always possible to obtain the NOQ state by restricting the parameters of the DSS.

However, the optimal parameters for the DSS found in Ref.~\cite{2010Choi} do not in general satisfy $\alpha_2 = -\alpha_1$ and $r_1 = r_2$, and this relationship between the displacements and squeeze parameters is true only for $\omega_q/\omega_c \gg 1$. Thus, only for $\omega_q/\omega_c \gg 1$ will the optimal NOQ state coincide with the optimal DSS. Interestingly, the we find that the optimal NOQ state is as good as the optimal DSS at capturing the nature of the true ground state in all parameter regimes examined, but based on physical intuition, the NOQ state requires two fewer free parameters.

The motivation behind having two distinct coherent state amplitudes in the DSS, $\alpha_1$ and $\alpha_2$, comes from a semiclassical treatment of the Rabi Hamiltonian, for which the effective potential has two distinct minimum for $g \simeq g^*$, corresponding to the two coherent state solutions $\alpha_1$ and $\alpha_2$ \cite{2010Choi}. However, as the effective potential has corrugations on the single photon level (due to the large coupling strength $g$), a rigorous derivation of any semiclassical model is not obviously apparent, and intuition gained from such an approach may not be valid.

The intuition for the single coherent state amplitude found in our NOQ state comes directly from a fully quantum treatment of the Rabi Hamiltonian in the large coupling limit, as discussed in section \ref{sec:limits}. Similarly, the necessity of a single, qubit-independent squeeze parameter can be derived from a perturbative treatment of the Rabi Hamiltonian at weak coupling, also discussed in section \ref{sec:limits}. These approaches are justified even when quantum effects in the interaction cannot be ignored, and the intuition gained from them can be used to create a variational state without concern for its validity. Indeed, the success of our NOQ state, despite its simple form, is testament to this fact.


\bibliography{References}

\end{document}